\documentclass[aps,prl,floatfix,twocolumn,showpacs,10pt]{revtex4-2}

\usepackage{graphicx}
\usepackage{dcolumn}
\usepackage{bm}
\usepackage{amssymb}
\usepackage{rotating} 
\usepackage{xcolor}

\begin{document}

\title{Electronic Excited States from a Variance-Based Contracted Quantum Eigensolver}

\author{Yuchen Wang and David A. Mazziotti}

\email{damazz@uchicago.edu}
\affiliation{Department of Chemistry and The James Franck Institute, The University of Chicago, Chicago, IL 60637}%

\date{Submitted May 6, 2023}


\begin{abstract}
Electronic excited states of molecules are central to many physical and chemical processes, and yet they are typically more difficult to compute than ground states.  In this paper we leverage the advantages of quantum computers to develop an algorithm for the highly accurate calculation of excited states.    We solve a contracted Schr{\"o}dinger equation (CSE)---a contraction (projection) of the Schr{\"o}dinger equation onto the space of two electrons---whose solutions correspond identically to the ground and excited states of the Schr{\"o}dinger equation.  While recent quantum algorithms for solving the CSE, known as contracted quantum eigensolvers (CQE), have focused on ground states, we develop a CQE based on the variance that is designed to optimize rapidly to a ground or excited state.  We apply the algorithm in a classical simulation without noise to computing the ground and excited states of H$_{4}$ and BH.
\end{abstract}


\maketitle

{\em Introduction:} Electronic excited states of molecules are critically important in any physical or chemical process that is not confined to the ground state such as photoabsorption and emission~\cite{Yam.2023}, non-adiabatic dynamics~\cite{Nelson.2022, Wang.202134w}, and electron scattering and transport~\cite{Greenwald.2021,Hsu.2017}.  Despite their central importance excited states are more difficult to compute than ground states~\cite{Gonzalez.2021, Benavides-Riveros.2022}.  Typical approaches compute the excited states as a response to the ground state~\cite{Monkhorst.1977,Runge.1984,Stanton.1993,Mazziotti.2003,Casida.2012}, which has limitations whenever excited states differ substantially from the ground state, e.g. in double-or multi-excitation processes~\cite{Maitra.2004}, charge-transfer states~\cite{Dreuw.2004,Mester.2022}, core excitations~\cite{Moitra.2023}, Rydberg states~\cite{Li.2022}, as well as conical intersections~\cite{Snyder.2011u3,Wang.2021si}.

One promising direction is to harness the potential advantages of quantum computers~\cite{Head-Marsden.2021,Bharti.2022}.  In the absence of noise quantum computers can prepare and measure quantum states whose wave functions are challenging to represent and manipulate on classical devices, potentially realizing significant advantages relative to classical devices~\cite{Lloyd.1993}.  While recent molecular algorithms have primarily focused on computing ground states~\cite{Bharti.2022} or obtaining multiple excited states at once from response theory~\cite{Gao.2021, Asthana.2022, Kumar.2022, Hlatshwayo.2022, Asthana.2023, Kim.2023} or a Krylov expansion~\cite{McClean.2017, Colless.2018, Nakanishi.2019, Bian.2019, Motta.2020, McClean.2020, Takeshita.2020, Gao.2021, Francis.2022, Huang.2022, Tkachenko.2022, Cortes.2022, Shen.2022, Yoshioka.2022, Motta.2023, Kanno.2023, Choi.2023}, quantum computers may be particularly well suited to realizing more accurate and direct calculations of excited states.  The possible advantages for ground states are in principle amplified for excited states.

In this paper we develop an algorithm for the highly accurate state-specific calculation of excited states on quantum devices.  Consider the contraction of the Schr{\"o}dinger equation onto the space of two electrons, known as the contracted Schr{\"o}dinger equation (CSE)~\cite{Mazziotti.1998e39, Nakatsuji.1996, Colmenero.1993, Mazziotti.20060v3, Boyn.2021}.  The CSE has two significant properties: (i) its solutions correspond identically to the ground-and excited-state solutions of the Schr{\"o}dinger equation~\cite{Mazziotti.1998e39, Nakatsuji.1996} and (ii) its compact structure reveals an exact two-body exponential parameterization of both ground and excited states~\cite{Mazziotti.20049l, Mazziotti.2020}.  Recent quantum-computing algorithms for solving the CSE or a part of the CSE, known as contracted quantum eigensolvers (CQEs)~\cite{Smart.2021, Boyn.2021u94, Smart.2022l2, Smart.2022w8u, Smart.2022, Smart.2023}, have mainly focused on the ground state.   We develop a CQE based on the energy variance that is designed to optimize rapidly to a ground or excited state.  To demonstrate, we apply the algorithm in a classical simulation without noise to computing the ground and excited states of H$_{4}$ and BH.

{\em Theory:} For a many-electron system consider the Schr{\"o}dinger equation
\begin{equation}
 ( {\hat H} - E_{n} ) | \Psi_{n} \rangle = 0
\end{equation}
in which ${\hat H}$ is the Hamiltonian operator and $| \Psi_{n} \rangle$ is the $N$-electron wave function for the $n^{\rm th}$ state. Contraction over all electrons except two generates the CSE~\cite{Mazziotti.1998e39, Nakatsuji.1996, Colmenero.1993, Mazziotti.20060v3, Boyn.2021, Smart.2021}
\begin{equation}
\langle \Psi_{n} | {\hat a}^{\dagger}_{i} {\hat a}^{\dagger}_{j} {\hat a}^{}_{l}  {\hat a}^{}_{k} ( {\hat H} - E_{n} )| \Psi_{n} \rangle = 0
\end{equation}
where ${\hat a}^{\dagger}_{i}$ and ${\hat a}_{i}$ are the creation and the annihilation operators for the $i^{\rm th}$ orbital.  As proved by Nakatsuji~\cite{Nakatsuji.1976} in first quantization and one of the authors~\cite{Mazziotti.1998e39} in second quantization, the CSE is satisfied by a wave function $| \Psi_{n} \rangle$ if and only if it satisfies the Schr{\"o}dinger equation.  The proofs show that the CSE implies the energy variance which implies the Schr{\"o}dinger equation.  Hence, the CSE determines a set of ground and excited states that is identical to that of the Schr{\"o}dinger equation.

As shown previously, the CSE can be solved for the ground-state wave function by minimizing the following energy functional iteratively on a quantum computer~\cite{Smart.2021, Boyn.2021u94, Smart.2022l2, Smart.2022w8u, Smart.2022, Smart.2023}
\begin{equation}
\min_{^{2} F_{m}}{ E[\Psi_{m}[^{2} F_{m}]]}
\end{equation}
where
\begin{equation}
\label{eq:cseansatz}
| \Psi_{m} \rangle = {\rm e}^{\hat F_{m}} | \Psi_{m-1} \rangle
\end{equation}
in which
\begin{equation}
{\hat F}_{m} = \sum_{pqst}{ ^{2} F^{pq;st}_{m} {\hat a}^{\dagger}_{p} {\hat a}^{\dagger}_{q} {\hat a}^{}_{t} {\hat a}^{}_{s} }
\end{equation}
This wave function is the CSE ansatz with the special property that its iterative minimization with respect to each two-body operator ${\hat F}_{m}$ converges to an exact solution of the CSE and hence, an exact solution of the Schr{\"o}dinger equation within a given finite basis set~\cite{Mazziotti.20049l, Mazziotti.2020}.  The gradient of the energy with respect to the latest $^{2} F_{m}$ is the residual of the CSE.  Hence, the gradient vanishes if and only if the CSE is satisfied.  We can also implement subsets of the CSE ansatz on a quantum computer.  For example, we have restricted the two-body operators ${\hat F_{m}}$ to be anti-Hermitian which generates strictly unitary transformations~\cite{Smart.2021, Boyn.2021u94, Smart.2022l2, Smart.2022w8u, Smart.2022}.  In this case the vanishing of the gradient causes the anti-Hermitian part of the CSE, known as the ACSE~\cite{Mazziotti.20060v3, Boyn.2021, Mazziotti.2007k2h, Mazziotti.2007, Gidofalvi.2009, Snyder.2010}, to be satisfied.

To extend to excited states, we replace the iterative minimization of the energy by an iterative minimization of the energy variance
\begin{equation}
\min_{^{2} F_{m}}{ {\rm Var}[\Psi_{m}[^{2} F_{m}]]}
\end{equation}
where
\begin{equation}
{\rm Var}[\Psi_{m}[^{2} F_{m}]] = \langle \Psi_{m} | ( {\hat H} - E_{m} )^{2} | \Psi_{m} \rangle
\end{equation}
in which
\begin{equation}
E_{m} = \langle \Psi_{m} | {\hat H} | \Psi_{m} \rangle
\end{equation}
with the wave function given by the CSE ansatz in Eq.~(\ref{eq:cseansatz}).  Throughout we assume that the wave function $| \Psi_{m} \rangle$ has been renormalized to one if necessary.  While the excited states are saddle points of the energy, they are minima of the variance. Moreover, any minimum is an exact stationary-state solution of the Schr{\"o}dinger equation (and the CSE) if the variance vanishes.  The variance has recently been applied for excited states in the context of the variational quantum eigensolver~\cite{Zhang.2021, Zhang.2022, Hobday.2022, Boyd.2022, Liu.2023}; however, in these studies the variance is only used to guide the optimization.  Here we use the CSE, which implies the variance~\cite{Nakatsuji.1976,Mazziotti.1998e39}, to not only perform the optimization but also to determine the iterative structure of the wave function in Eq.~(\ref{eq:cseansatz}).  The CSE ansatz is formally exact with the important property that it remains exact even without reoptimization of the $^{2} F_{m-q}$ for $q>0$ from previous iterations.  The gradient of the variance with respect to $^{2} F_{m}$ can be computed as follows:
\begin{equation}
\label{eq:grad}
\frac{ \partial {\rm Var} }{ \partial \left ( ^{2} F^{st;pq}_{m} \right ) } = 2 \langle {\Psi_{m}} | ( {\hat \Gamma}^{pq}_{st} - {}^{2} D^{pq}_{st} ) ( {\hat H} - E_{m} )^{2} | {\Psi_{m}} \rangle ,
\end{equation}
in which ${\hat \Gamma}^{pq}_{st} = {\hat a}^{\dagger}_{p} {\hat a}^{\dagger}_{q} {\hat a}^{}_{t} {\hat a}^{}_{s}$ and the elements of the 2-RDM are
\begin{equation}
^{2} D^{pq}_{st} = \langle \Psi_{m} | {\hat \Gamma}^{pq}_{st} | \Psi_{m} \rangle.
\end{equation}
Practically, we can approximate the minimization of the variance at the $m^{\rm th}$ iteration by selecting $^{2} F_{m}$ to be proportional to the direction of the gradient or a related search direction from any gradient-descent method with the proportionality constant (or step size) being determined by a line search.  Other related generalizations of the variational principle in the CQE can also be considered. For example, we can: (1) solve the CSE or ACSE directly for the wave function, (2) minimize the least-squares norm of the CSE or ACSE, or (3) augment the variance functional with an additional functional such as a small amount of the energy functional.

Optimizing the energy variance is ideal for a quantum computer.  While computing the variance requires not only the two-particle reduced density matrix (2-RDM) but also the four-particle RDM on a classical computer, we can readily compute it at the $m^{\rm th}$ iteration on a quantum computer by introducing an ancillary qubit to generate an extra wave function
\begin{equation}
| {\tilde \Psi_{m}} \rangle = {\rm e}^{i \delta \left ( {\hat H} - E_{m} \right ) } | \Psi_{m} \rangle
\end{equation}
such that
\begin{equation}
\label{eq:mvar}
\langle \Psi_{m} | ( {\hat H} - E_{m} )^{2} | \Psi_{m} \rangle \approx \frac{ 1 - \Re \langle \Psi_{m} | {\tilde \Psi}_{m} \rangle}{\delta^{2}/2}
\end{equation}
where $\Re(z)$ returns the real part of $z$, the approximation is accurate to $O(\delta^{2})$, and $\delta$ is a small parameter. This formula is an extension of the difference formulas employed in previous CQE algorithms~\cite{Smart.2021} as well as in the context of open quantum systems~\cite{Schlimgen.2021}.  As shown in previous work, the limit of $\delta$ approaching zero can be computed by using Richardson’s extrapolation from a series of $\delta$ values~\cite{Schlimgen.2021, Seki.2021}.  Recently we have shown how the residuals of both the CSE and ACSE can be efficiently calculated on a quantum computer from only a 2-RDM-like tomography~\cite{Smart.2021, Smart.2023}.  Similarly, the key term in the gradient of the variance with respect to $^{2} F$ in the CSE wave-function ansatz can be computed from a 2-RDM-like tomography
\begin{equation}
\label{eq:mgrad}
\langle \Psi_{m} | {\hat \Gamma}^{pq}_{st} ( {\hat H} - E_{m} )^{2} | \Psi_{m} \rangle \approx \frac{{}^{2} D^{pq}_{st} - \Re \langle \Psi_{m} | {\hat \Gamma}^{pq}_{st}  | \tilde \Psi_{m}\rangle}{\delta^2/2}
\end{equation}
where the approximation is accurate to $O(\delta^{2})$.  While the left side formally depends upon the six-particle RDM, through a combination of state preparation and tomography, we can obtain the gradient of the variance with the CSE ansatz from only the measurement of the two-particle reduced transition matrix between the states $| \Psi_{m} \rangle$ and $| {\tilde \Psi}_{m} \rangle$.  Formulas in Eqs.~(\ref{eq:mvar}) and~(\ref{eq:mgrad}) assume that the Hamiltonian and wave function are real, but as in Refs.~\cite{Smart.2021,Schlimgen.2021}, they can be readily generalized through additional measurements to treat complex Hamiltonians and wave functions as well as to realize higher-order approximations.  The algorithm for the variance-based CQE for excited states is summarized in Table~\ref{tab:alg}.

\begin{table}[t!]
  \caption{\normalsize Variance-based CQE algorithm.}
  \label{tab:alg}
  \begin{ruledtabular}
  \begin{tabular}{l}
  {\bf Algorithm: Variance-based CQE} \\
  \hspace{0.0in} {\rm Given} $m=0$ {\rm and convergence tolerance} $\epsilon$. \\
  \hspace{0.0in} {\rm Choose initial wave function} $| \Psi_{0} \rangle$. \\
  \hspace{0.0in} {\rm Repeat until the energy variance is less than} $\epsilon$. \\
  \hspace{0.1in} {\rm {\bf Step 1:} Prepare} $| \tilde \Psi_{m} \rangle = {\rm e}^{i \delta ( {\hat H} - E_{m} )} | \Psi_{m} \rangle$  \\
  \hspace{0.1in} {\rm {\bf Step 2:} Measure variance using Eq.~(\ref{eq:mvar})} \\
  \hspace{0.1in} {\rm {\bf Step 3:} Measure} $\langle \Psi_{m} | {\hat a}^{\dagger}_{p} {\hat a}^{\dagger}_{q} {\hat a}^{}_{t} {\hat a}^{}_{s}  | {\tilde \Psi}_{m} \rangle$ in Eq.~(\ref{eq:mgrad}) \\
  \hspace{0.1in} {\rm {\bf Step 4:} Compute gradient from Eqs.~(\ref{eq:grad}) and~(\ref{eq:mgrad})} \\
  \hspace{0.1in} {\rm {\bf Step 5:} Compute gradient-descent search direction} ${\hat F}_{m+1}$ \\
  \hspace{0.1in} {\rm {\bf Step 6:} Prepare} $| \Psi_{m+1} \rangle = {\rm e}^{{\hat F}_{m+1}} | \Psi_{m} \rangle$  \\
  \hspace{0.1in} {\rm {\bf Step 7:} Optimize magnitude of ${\hat F}_{m+1}$ via Steps~1, 2, and~6} \\
  \hspace{0.1in} {\rm {\bf Step 8:} Set} $m = m+1$.
  \end{tabular}
  \end{ruledtabular}
\end{table}

\begin{table*}[t!]
  \caption{\normalsize The energy, energy error, variance, and least-squares CSE norm of the ground state and each of the first 15~excited states of linear H$_{4}$ from the variance-based CQE are shown.  Energies are given in hartrees.}
  \label{tab:h4}
  \begin{ruledtabular}
  \begin{tabular}{cccccccc}
\multicolumn{1}{c}{State} & \multicolumn{1}{c}{$2S+1$}	 & \multicolumn{1}{c}{$\langle {\hat S}_{z} \rangle$} & \multicolumn{1}{c}{Energy} & \multicolumn{1}{c}{Iterations}	& \multicolumn{1}{c}{Energy Error} & \multicolumn{1}{c}{Variance} & \multicolumn{1}{c}{CSE Norm} \\ \hline
0 & 1 &~0 & -2.18096635 & 20 & $6.6 \times 10^{-7}$ & $8.0 \times 10^{-7}$  & $4.3 \times 10^{-8}$ \\
1 & 3 &-1 & -1.95019128 &  8 & $4.0 \times 10^{-7}$ & $4.1 \times 10^{-7}$  & $3.4 \times 10^{-8}$ \\
2 & 3 & ~0 & -1.95019128 &   7 & $3.9 \times 10^{-7}$ & $5.2 \times 10^{-7}$  & $2.3 \times 10^{-8}$ \\	
3 & 3 & ~1 & -1.95019128 &   8 & $4.0 \times 10^{-7}$ & $4.1 \times 10^{-7}$  & $3.4 \times 10^{-8}$ \\
4 & 3 & -1 & -1.73654709 &  13 & $6.4 \times 10^{-7}$ & $3.7 \times 10^{-7}$  & $3.9 \times 10^{-8}$ \\
5 & 3 & ~0 & -1.73654709 &   9 & $1.7 \times 10^{-6}$ & $7.8 \times 10^{-7}$  & $3.9 \times 10^{-8}$ \\
6 & 3 & ~1 & -1.73654709 &  13 & $6.4 \times 10^{-6}$ & $3.7 \times 10^{-7}$  & $3.9 \times 10^{-8}$ \\
7 & 1 & ~0 & -1.66711149 &  17 & $8.6 \times 10^{-7}$ & $9.8 \times 10^{-7}$  & $6.5 \times 10^{-8}$ \\
8 & 1 & ~0 & -1.63892672 &   9 & $4.1 \times 10^{-7}$ & $3.3 \times 10^{-7}$  & $2.0 \times 10^{-8}$ \\
9 & 3 & -1 & -1.45713456 &  17 & $7.9 \times 10^{-7}$ & $6.0 \times 10^{-7}$  & $7.1 \times 10^{-8}$ \\
10&3 &~0 & -1.45713456 &  21 & $7.7 \times 10^{-8}$ & $9.5 \times 10^{-7}$  & $7.6 \times 10^{-8}$ \\
11&3 &~1 & -1.45713456 &  17 & $7.9 \times 10^{-7}$ & $6.0 \times 10^{-7}$  & $7.1 \times 10^{-8}$ \\
12&1 &~0 & -1.34940191 &  37 & $9.1 \times 10^{-7}$ & $8.6 \times 10^{-7}$  & $5.4 \times 10^{-8}$ \\
13&3 & -1 & -1.30398471 &  37 & $9.8 \times 10^{-6}$ & $7.3 \times 10^{-7}$  & $6.4 \times 10^{-8}$ \\
14&3 &~0 & -1.30398471 &  11 & $2.8 \times 10^{-7}$ & $3.7 \times 10^{-7}$  & $2.2 \times 10^{-8}$ \\
15&3 &~1 & -1.30398471 &  39 & $1.4 \times 10^{-5}$ & $9.6 \times 10^{-7}$  & $1.0 \times 10^{-7}$
  \end{tabular}
  \end{ruledtabular}
\end{table*}

{\em Results:} To demonstrate, we apply the variance-based CQE algorithm in a classical simulation without noise to computing the excited states of the molecules H$_{4}$ and BH.  The H$_{4}$ molecule is treated in its linear conformation with adjacent hydrogen atoms separated by 1~\AA.  We use a minimal Slater-type orbital (STO-6G) basis set~\cite{Hehre.1969} for both molecules as well as a frozen 1s core for the boron atom in BH.  Molecular orbitals from the Hartree-Fock method and one-and two-electron integrals are obtained with the Quantum Chemistry Package in Maple~\cite{rdmchem}.  In implementing the algorithm in Table~\ref{tab:alg}, we restrict the ${\hat F}$ operators to be anti-Hermitian, making the two-body exponential transformations unitary and perform exact line searches along directions from a limited-memory Broyden–Fletcher–Goldfarb–Shanno (BFGS) method~\cite{Liu.1989} with a single gradient stored.  Initial guesses for the wave function are the Slater determinants from the Hartree-Fock orbitals; when $\langle {\hat S}_{z} \rangle = \pm 1$, we use a single high-spin Slater determinant, but when $\langle {\hat S}_{z} \rangle = 0$, unless noted otherwise, we use an equal linear combination of two determinants that are related by switching the $\alpha$ (spin up) and $\beta$ (spin down) orbitals with the relative phases being +1 for a singlet and -1 for a triplet.

\begin{figure}[tb!]
    \centering
    \includegraphics[scale=0.4]{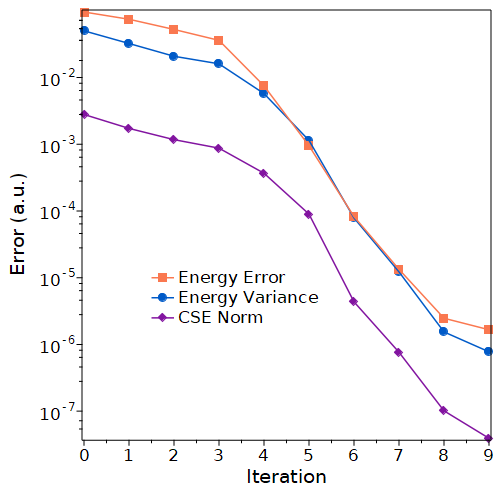}
    \caption{Superlinear convergence of the energy error, variance, and least-squares CSE norm is shown for the fifth excited state of linear H$_{4}$.}
    \label{fig:h4}
\end{figure}

The ground state and the first 15 excited states of linear H$_{4}$ as computed from the variance-based CQE are shown in Table~\ref{tab:h4}.  The algorithm is performed iteratively until the energy variance is less than $10^{-6}$~a.u.  The number of iterations required for convergence varies from 7 for the second excited state to 39 for the fifteen excited state.  At convergence the energy error is also less than $10^{-6}$~hartrees except for the fifth, thirteenth, and fifteenth excited states.  Even though the energies of the excited states need not be upper bounds to the energies from exact diagonalization, we find that all excited-state energies are strictly above those from diagonalization.  We also compute the least-squares error in the CSE---the sum of the squares of the errors in the CSE, which is approximately an order of magnitude less than the energy variance for each state.  For the fifth excited state Fig.~\ref{fig:h4} shows the convergence of the energy error, variance, and least-squares CSE norm.  We observe superlinear convergence towards zero in all three metrics for the error.

\begin{figure}[tb!]
    \centering
    \includegraphics[scale=0.4]{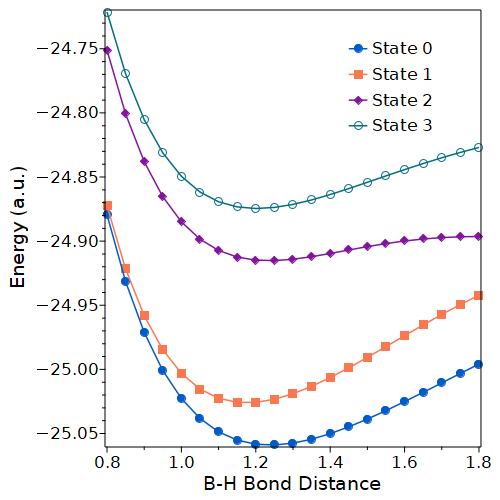}
    \caption{Energies of the ground state and the first three excited states of BH are shown as functions of the bond distance.  Symbols represent variance-based CQE energies while solid lines represent energies from exact diagonalization.}
    \label{fig:bh}
\end{figure}

The energies of the ground state and the first three excited states of BH are shown as functions of the bond distance in Fig.~\ref{fig:bh}.  The solid lines denote the ground- and excited-state energies from exact diagonalization while the symbols denote the energies from the variance-based CQE. In each case the energy variance in the CQE is converged to less than $10^{-5}$~a.u. We observe that the CQE reproduces the potential energy curves with maximum energy errors of 0.00001, 0.00008, 0.00004, and 0.00024~hartrees for the ground and first three excited states, respectively.

{\em Conclusions:} Here we present a variance-based CQE for computing highly accurate molecular excited states on quantum computers.  The CQE is a family of algorithms in which a contraction of the Schr{\"o}dinger equation to the space of two particles (CSE) is solved for stationary-state energies and their 2-RDMs.   The structure of the CSE implies an exact ansatz for any ground- or excited-state wave function in which a two-body exponential transformation is iteratively applied and optimized to update a trial wave function.  Importantly, unlike iterative variational quantum eigensolvers, the CQE does not need to reoptimize previous transformations to satisfy the CSE and thereby solve the Schr{\"o}dinger equation.  While recent work with CQE has focused on the ground state, here we present a CQE algorithm for excited states in which we iteratively minimize the energy variance with respect to the CSE (or ACSE) ansatz.  We show that the variance-based CQE yields highly accurate ground-and excited-state energies for the example cases of H$_{4}$ and BH in the absence of noise.  Future work will examine the application of the variance-based CQE on noisy intermediate-scale quantum (NISQ) computers. The present approach represents an important step towards the accurate modeling of molecular excited states on NISQ and fault-tolerant quantum computers.

\begin{acknowledgments}

D.A.M. gratefully acknowledges the Department of Energy, Office of Basic Energy Sciences Grant DE-SC0019215, the U.S. National Science Foundation Grants CHE-2155082 and No. CHE-2035876.

\end{acknowledgments}

\bibliography{EX-CQE-BIB}

\begin{thebibliography}{72}%
\makeatletter
\providecommand \@ifxundefined [1]{%
 \@ifx{#1\undefined}
}%
\providecommand \@ifnum [1]{%
 \ifnum #1\expandafter \@firstoftwo
 \else \expandafter \@secondoftwo
 \fi
}%
\providecommand \@ifx [1]{%
 \ifx #1\expandafter \@firstoftwo
 \else \expandafter \@secondoftwo
 \fi
}%
\providecommand \natexlab [1]{#1}%
\providecommand \enquote  [1]{``#1''}%
\providecommand \bibnamefont  [1]{#1}%
\providecommand \bibfnamefont [1]{#1}%
\providecommand \citenamefont [1]{#1}%
\providecommand \href@noop [0]{\@secondoftwo}%
\providecommand \href [0]{\begingroup \@sanitize@url \@href}%
\providecommand \@href[1]{\@@startlink{#1}\@@href}%
\providecommand \@@href[1]{\endgroup#1\@@endlink}%
\providecommand \@sanitize@url [0]{\catcode `\\12\catcode `\$12\catcode
  `\&12\catcode `\#12\catcode `\^12\catcode `\_12\catcode `\%12\relax}%
\providecommand \@@startlink[1]{}%
\providecommand \@@endlink[0]{}%
\providecommand \url  [0]{\begingroup\@sanitize@url \@url }%
\providecommand \@url [1]{\endgroup\@href {#1}{\urlprefix }}%
\providecommand \urlprefix  [0]{URL }%
\providecommand \Eprint [0]{\href }%
\providecommand \doibase [0]{https://doi.org/}%
\providecommand \selectlanguage [0]{\@gobble}%
\providecommand \bibinfo  [0]{\@secondoftwo}%
\providecommand \bibfield  [0]{\@secondoftwo}%
\providecommand \translation [1]{[#1]}%
\providecommand \BibitemOpen [0]{}%
\providecommand \bibitemStop [0]{}%
\providecommand \bibitemNoStop [0]{.\EOS\space}%
\providecommand \EOS [0]{\spacefactor3000\relax}%
\providecommand \BibitemShut  [1]{\csname bibitem#1\endcsname}%
\let\auto@bib@innerbib\@empty
\bibitem [{\citenamefont {Yam}(2023)}]{Yam.2023}%
  \BibitemOpen
  \bibfield  {author} {\bibinfo {author} {\bibfnamefont {V.~W.-W.}\
  \bibnamefont {Yam}},\ }\bibfield  {title} {\bibinfo {title} {{Using synthesis
  to steer excited states and their properties and functions}},\ }\href
  {https://doi.org/10.1038/s44160-022-00202-5} {\bibfield  {journal} {\bibinfo
  {journal} {Nature Synthesis}\ }\textbf {\bibinfo {volume} {2}},\ \bibinfo
  {pages} {94} (\bibinfo {year} {2023})}\BibitemShut {NoStop}%
\bibitem [{\citenamefont {Nelson}\ \emph {et~al.}(2022)\citenamefont {Nelson},
  \citenamefont {Fernandez-Alberti},\ and\ \citenamefont
  {Tretiak}}]{Nelson.2022}%
  \BibitemOpen
  \bibfield  {author} {\bibinfo {author} {\bibfnamefont {T.~R.}\ \bibnamefont
  {Nelson}}, \bibinfo {author} {\bibfnamefont {S.}~\bibnamefont
  {Fernandez-Alberti}},\ and\ \bibinfo {author} {\bibfnamefont
  {S.}~\bibnamefont {Tretiak}},\ }\bibfield  {title} {\bibinfo {title}
  {{Modeling excited-state molecular dynamics beyond the Born–Oppenheimer
  regime}},\ }\href {https://doi.org/10.1038/s43588-022-00357-3} {\bibfield
  {journal} {\bibinfo  {journal} {Nature Computational Science}\ }\textbf
  {\bibinfo {volume} {2}},\ \bibinfo {pages} {689} (\bibinfo {year}
  {2022})}\BibitemShut {NoStop}%
\bibitem [{\citenamefont {Wang}\ \emph {et~al.}(2021)\citenamefont {Wang},
  \citenamefont {Guan}, \citenamefont {Guo},\ and\ \citenamefont
  {Yarkony}}]{Wang.202134w}%
  \BibitemOpen
  \bibfield  {author} {\bibinfo {author} {\bibfnamefont {Y.}~\bibnamefont
  {Wang}}, \bibinfo {author} {\bibfnamefont {Y.}~\bibnamefont {Guan}}, \bibinfo
  {author} {\bibfnamefont {H.}~\bibnamefont {Guo}},\ and\ \bibinfo {author}
  {\bibfnamefont {D.~R.}\ \bibnamefont {Yarkony}},\ }\bibfield  {title}
  {\bibinfo {title} {{Enabling complete multichannel nonadiabatic dynamics: A
  global representation of the two-channel coupled, 1,21A and 13A states of NH3
  using neural networks}},\ }\href {https://doi.org/10.1063/5.0037684}
  {\bibfield  {journal} {\bibinfo  {journal} {The Journal of Chemical Physics}\
  }\textbf {\bibinfo {volume} {154}},\ \bibinfo {pages} {094121} (\bibinfo
  {year} {2021})}\BibitemShut {NoStop}%
\bibitem [{\citenamefont {Greenwald}\ \emph {et~al.}(2021)\citenamefont
  {Greenwald}, \citenamefont {Cameron}, \citenamefont {Findlay}, \citenamefont
  {Fu}, \citenamefont {Gunasekaran}, \citenamefont {Skabara},\ and\
  \citenamefont {Venkataraman}}]{Greenwald.2021}%
  \BibitemOpen
  \bibfield  {author} {\bibinfo {author} {\bibfnamefont {J.~E.}\ \bibnamefont
  {Greenwald}}, \bibinfo {author} {\bibfnamefont {J.}~\bibnamefont {Cameron}},
  \bibinfo {author} {\bibfnamefont {N.~J.}\ \bibnamefont {Findlay}}, \bibinfo
  {author} {\bibfnamefont {T.}~\bibnamefont {Fu}}, \bibinfo {author}
  {\bibfnamefont {S.}~\bibnamefont {Gunasekaran}}, \bibinfo {author}
  {\bibfnamefont {P.~J.}\ \bibnamefont {Skabara}},\ and\ \bibinfo {author}
  {\bibfnamefont {L.}~\bibnamefont {Venkataraman}},\ }\bibfield  {title}
  {\bibinfo {title} {{Highly nonlinear transport across single-molecule
  junctions via destructive quantum interference}},\ }\href
  {https://doi.org/10.1038/s41565-020-00807-x} {\bibfield  {journal} {\bibinfo
  {journal} {Nature Nanotechnology}\ }\textbf {\bibinfo {volume} {16}},\
  \bibinfo {pages} {313} (\bibinfo {year} {2021})}\BibitemShut {NoStop}%
\bibitem [{\citenamefont {Hsu}\ \emph {et~al.}(2017)\citenamefont {Hsu},
  \citenamefont {Jin}, \citenamefont {Chen},\ and\ \citenamefont
  {Peng}}]{Hsu.2017}%
  \BibitemOpen
  \bibfield  {author} {\bibinfo {author} {\bibfnamefont {L.-Y.}\ \bibnamefont
  {Hsu}}, \bibinfo {author} {\bibfnamefont {B.-Y.}\ \bibnamefont {Jin}},
  \bibinfo {author} {\bibfnamefont {C.-h.}\ \bibnamefont {Chen}},\ and\
  \bibinfo {author} {\bibfnamefont {S.-M.}\ \bibnamefont {Peng}},\ }\bibfield
  {title} {\bibinfo {title} {{Reaction: New Insights into Molecular
  Electronics}},\ }\href {https://doi.org/10.1016/j.chempr.2017.08.007}
  {\bibfield  {journal} {\bibinfo  {journal} {Chem}\ }\textbf {\bibinfo
  {volume} {3}},\ \bibinfo {pages} {378} (\bibinfo {year} {2017})}\BibitemShut
  {NoStop}%
\bibitem [{\citenamefont {González}\ and\ \citenamefont
  {Lindh}(2021)}]{Gonzalez.2021}%
  \BibitemOpen
  \bibinfo {editor} {\bibfnamefont {L.}~\bibnamefont {González}}\ and\
  \bibinfo {editor} {\bibfnamefont {R.}~\bibnamefont {Lindh}},\ eds.,\ \href
  {https://doi.org/10.1002/9781119417774} {\emph {\bibinfo {title} {{Quantum
  Chemistry and Dynamics of Excited States}}}}\ (\bibinfo  {publisher}
  {Wiley},\ \bibinfo {address} {New York},\ \bibinfo {year} {2021})\BibitemShut
  {NoStop}%
\bibitem [{\citenamefont {Benavides-Riveros}\ \emph {et~al.}(2022)\citenamefont
  {Benavides-Riveros}, \citenamefont {Chen}, \citenamefont {Schilling},
  \citenamefont {Mantilla},\ and\ \citenamefont
  {Pittalis}}]{Benavides-Riveros.2022}%
  \BibitemOpen
  \bibfield  {author} {\bibinfo {author} {\bibfnamefont {C.~L.}\ \bibnamefont
  {Benavides-Riveros}}, \bibinfo {author} {\bibfnamefont {L.}~\bibnamefont
  {Chen}}, \bibinfo {author} {\bibfnamefont {C.}~\bibnamefont {Schilling}},
  \bibinfo {author} {\bibfnamefont {S.}~\bibnamefont {Mantilla}},\ and\
  \bibinfo {author} {\bibfnamefont {S.}~\bibnamefont {Pittalis}},\ }\bibfield
  {title} {\bibinfo {title} {{Excitations of Quantum Many-Body Systems via
  Purified Ensembles: A Unitary-Coupled-Cluster-Based Approach}},\ }\href
  {https://doi.org/10.1103/physrevlett.129.066401} {\bibfield  {journal}
  {\bibinfo  {journal} {Physical Review Letters}\ }\textbf {\bibinfo {volume}
  {129}},\ \bibinfo {pages} {066401} (\bibinfo {year} {2022})},\ \Eprint
  {https://arxiv.org/abs/2201.10974} {2201.10974} \BibitemShut {NoStop}%
\bibitem [{\citenamefont {Monkhorst}(1977)}]{Monkhorst.1977}%
  \BibitemOpen
  \bibfield  {author} {\bibinfo {author} {\bibfnamefont {H.~J.}\ \bibnamefont
  {Monkhorst}},\ }\bibfield  {title} {\bibinfo {title} {{Calculation of
  properties with the coupled‐cluster method}},\ }\href
  {https://doi.org/10.1002/qua.560120850} {\bibfield  {journal} {\bibinfo
  {journal} {International Journal of Quantum Chemistry}\ }\textbf {\bibinfo
  {volume} {12}},\ \bibinfo {pages} {421} (\bibinfo {year} {1977})}\BibitemShut
  {NoStop}%
\bibitem [{\citenamefont {Runge}\ and\ \citenamefont
  {Gross}(1984)}]{Runge.1984}%
  \BibitemOpen
  \bibfield  {author} {\bibinfo {author} {\bibfnamefont {E.}~\bibnamefont
  {Runge}}\ and\ \bibinfo {author} {\bibfnamefont {E.~K.~U.}\ \bibnamefont
  {Gross}},\ }\bibfield  {title} {\bibinfo {title} {{Density-Functional Theory
  for Time-Dependent Systems}},\ }\href
  {https://doi.org/10.1103/physrevlett.52.997} {\bibfield  {journal} {\bibinfo
  {journal} {Physical Review Letters}\ }\textbf {\bibinfo {volume} {52}},\
  \bibinfo {pages} {997} (\bibinfo {year} {1984})}\BibitemShut {NoStop}%
\bibitem [{\citenamefont {Stanton}\ and\ \citenamefont
  {Bartlett}(1993)}]{Stanton.1993}%
  \BibitemOpen
  \bibfield  {author} {\bibinfo {author} {\bibfnamefont {J.~F.}\ \bibnamefont
  {Stanton}}\ and\ \bibinfo {author} {\bibfnamefont {R.~J.}\ \bibnamefont
  {Bartlett}},\ }\bibfield  {title} {\bibinfo {title} {{The equation of motion
  coupled‐cluster method. A systematic biorthogonal approach to molecular
  excitation energies, transition probabilities, and excited state
  properties}},\ }\href {https://doi.org/10.1063/1.464746} {\bibfield
  {journal} {\bibinfo  {journal} {The Journal of Chemical Physics}\ }\textbf
  {\bibinfo {volume} {98}},\ \bibinfo {pages} {7029} (\bibinfo {year}
  {1993})}\BibitemShut {NoStop}%
\bibitem [{\citenamefont {Mazziotti}(2003)}]{Mazziotti.2003}%
  \BibitemOpen
  \bibfield  {author} {\bibinfo {author} {\bibfnamefont {D.~A.}\ \bibnamefont
  {Mazziotti}},\ }\bibfield  {title} {\bibinfo {title} {{Extraction of
  electronic excited states from the ground-state two-particle reduced density
  matrix}},\ }\href {https://doi.org/10.1103/physreva.68.052501} {\bibfield
  {journal} {\bibinfo  {journal} {Physical Review A}\ }\textbf {\bibinfo
  {volume} {68}},\ \bibinfo {pages} {052501} (\bibinfo {year}
  {2003})}\BibitemShut {NoStop}%
\bibitem [{\citenamefont {Casida}\ and\ \citenamefont
  {Huix-Rotllant}(2012)}]{Casida.2012}%
  \BibitemOpen
  \bibfield  {author} {\bibinfo {author} {\bibfnamefont {M.}~\bibnamefont
  {Casida}}\ and\ \bibinfo {author} {\bibfnamefont {M.}~\bibnamefont
  {Huix-Rotllant}},\ }\bibfield  {title} {\bibinfo {title} {{Progress in
  Time-Dependent Density-Functional Theory}},\ }\href
  {https://doi.org/10.1146/annurev-physchem-032511-143803} {\bibfield
  {journal} {\bibinfo  {journal} {Annual Review of Physical Chemistry}\
  }\textbf {\bibinfo {volume} {63}},\ \bibinfo {pages} {287} (\bibinfo {year}
  {2012})},\ \Eprint {https://arxiv.org/abs/1108.0611} {1108.0611} \BibitemShut
  {NoStop}%
\bibitem [{\citenamefont {Maitra}\ \emph {et~al.}(2004)\citenamefont {Maitra},
  \citenamefont {Zhang}, \citenamefont {Cave},\ and\ \citenamefont
  {Burke}}]{Maitra.2004}%
  \BibitemOpen
  \bibfield  {author} {\bibinfo {author} {\bibfnamefont {N.~T.}\ \bibnamefont
  {Maitra}}, \bibinfo {author} {\bibfnamefont {F.}~\bibnamefont {Zhang}},
  \bibinfo {author} {\bibfnamefont {R.~J.}\ \bibnamefont {Cave}},\ and\
  \bibinfo {author} {\bibfnamefont {K.}~\bibnamefont {Burke}},\ }\bibfield
  {title} {\bibinfo {title} {{Double excitations within time-dependent density
  functional theory linear response}},\ }\href
  {https://doi.org/10.1063/1.1651060} {\bibfield  {journal} {\bibinfo
  {journal} {The Journal of Chemical Physics}\ }\textbf {\bibinfo {volume}
  {120}},\ \bibinfo {pages} {5932} (\bibinfo {year} {2004})}\BibitemShut
  {NoStop}%
\bibitem [{\citenamefont {Dreuw}\ and\ \citenamefont
  {Head-Gordon}(2004)}]{Dreuw.2004}%
  \BibitemOpen
  \bibfield  {author} {\bibinfo {author} {\bibfnamefont {A.}~\bibnamefont
  {Dreuw}}\ and\ \bibinfo {author} {\bibfnamefont {M.}~\bibnamefont
  {Head-Gordon}},\ }\bibfield  {title} {\bibinfo {title} {{Failure of
  Time-Dependent Density Functional Theory for Long-Range Charge-Transfer
  Excited States: The Zincbacteriochlorin-Bacteriochlorin and
  Bacteriochlorophyll-Spheroidene Complexes}},\ }\href
  {https://doi.org/10.1021/ja039556n} {\bibfield  {journal} {\bibinfo
  {journal} {Journal of the American Chemical Society}\ }\textbf {\bibinfo
  {volume} {126}},\ \bibinfo {pages} {4007} (\bibinfo {year}
  {2004})}\BibitemShut {NoStop}%
\bibitem [{\citenamefont {Mester}\ and\ \citenamefont
  {K{\'a}llay}(2022)}]{Mester.2022}%
  \BibitemOpen
  \bibfield  {author} {\bibinfo {author} {\bibfnamefont {D.}~\bibnamefont
  {Mester}}\ and\ \bibinfo {author} {\bibfnamefont {M.}~\bibnamefont
  {K{\'a}llay}},\ }\bibfield  {title} {\bibinfo {title} {{Charge-Transfer
  Excitations within Density Functional Theory: How Accurate Are the Most
  Recommended Approaches?}},\ }\href {https://doi.org/10.1021/acs.jctc.1c01307}
  {\bibfield  {journal} {\bibinfo  {journal} {Journal of Chemical Theory and
  Computation}\ }\textbf {\bibinfo {volume} {18}},\ \bibinfo {pages} {1646}
  (\bibinfo {year} {2022})}\BibitemShut {NoStop}%
\bibitem [{\citenamefont {Moitra}\ \emph {et~al.}(2023)\citenamefont {Moitra},
  \citenamefont {Konecny}, \citenamefont {Kadek}, \citenamefont {Rubio},\ and\
  \citenamefont {Repisky}}]{Moitra.2023}%
  \BibitemOpen
  \bibfield  {author} {\bibinfo {author} {\bibfnamefont {T.}~\bibnamefont
  {Moitra}}, \bibinfo {author} {\bibfnamefont {L.}~\bibnamefont {Konecny}},
  \bibinfo {author} {\bibfnamefont {M.}~\bibnamefont {Kadek}}, \bibinfo
  {author} {\bibfnamefont {A.}~\bibnamefont {Rubio}},\ and\ \bibinfo {author}
  {\bibfnamefont {M.}~\bibnamefont {Repisky}},\ }\bibfield  {title} {\bibinfo
  {title} {{Accurate Relativistic Real-Time Time-Dependent Density Functional
  Theory for Valence and Core Attosecond Transient Absorption Spectroscopy.}},\
  }\href {https://doi.org/10.1021/acs.jpclett.2c03599} {\bibfield  {journal}
  {\bibinfo  {journal} {The journal of physical chemistry letters}\ }\textbf
  {\bibinfo {volume} {14}},\ \bibinfo {pages} {1714} (\bibinfo {year}
  {2023})},\ \Eprint {https://arxiv.org/abs/2211.16383} {2211.16383}
  \BibitemShut {NoStop}%
\bibitem [{\citenamefont {Li}\ \emph {et~al.}(2022)\citenamefont {Li},
  \citenamefont {Jin}, \citenamefont {Su},\ and\ \citenamefont
  {Yang}}]{Li.2022}%
  \BibitemOpen
  \bibfield  {author} {\bibinfo {author} {\bibfnamefont {J.}~\bibnamefont
  {Li}}, \bibinfo {author} {\bibfnamefont {Y.}~\bibnamefont {Jin}}, \bibinfo
  {author} {\bibfnamefont {N.~Q.}\ \bibnamefont {Su}},\ and\ \bibinfo {author}
  {\bibfnamefont {W.}~\bibnamefont {Yang}},\ }\bibfield  {title} {\bibinfo
  {title} {{Combining localized orbital scaling correction and Bethe-Salpeter
  equation for accurate excitation energies.}},\ }\href
  {https://doi.org/10.1063/5.0087498} {\bibfield  {journal} {\bibinfo
  {journal} {The Journal of chemical physics}\ }\textbf {\bibinfo {volume}
  {156}},\ \bibinfo {pages} {154101} (\bibinfo {year} {2022})},\ \Eprint
  {https://arxiv.org/abs/2207.00508} {2207.00508} \BibitemShut {NoStop}%
\bibitem [{\citenamefont {Snyder}\ and\ \citenamefont
  {Mazziotti}(2011)}]{Snyder.2011u3}%
  \BibitemOpen
  \bibfield  {author} {\bibinfo {author} {\bibfnamefont {J.~W.}\ \bibnamefont
  {Snyder}}\ and\ \bibinfo {author} {\bibfnamefont {D.~A.}\ \bibnamefont
  {Mazziotti}},\ }\bibfield  {title} {\bibinfo {title} {{Photoexcited
  conversion of gauche-1,3-butadiene to bicyclobutane via a conical
  intersection: Energies and reduced density matrices from the anti-Hermitian
  contracted Schrödinger equation}},\ }\href
  {https://doi.org/10.1063/1.3606466} {\bibfield  {journal} {\bibinfo
  {journal} {The Journal of Chemical Physics}\ }\textbf {\bibinfo {volume}
  {135}},\ \bibinfo {pages} {024107} (\bibinfo {year} {2011})}\BibitemShut
  {NoStop}%
\bibitem [{\citenamefont {Wang}\ and\ \citenamefont
  {Yarkony}(2021)}]{Wang.2021si}%
  \BibitemOpen
  \bibfield  {author} {\bibinfo {author} {\bibfnamefont {Y.}~\bibnamefont
  {Wang}}\ and\ \bibinfo {author} {\bibfnamefont {D.~R.}\ \bibnamefont
  {Yarkony}},\ }\bibfield  {title} {\bibinfo {title} {{Conical intersection
  seams in spin–orbit coupled systems with an even number of electrons: A
  numerical study based on neural network fit surfaces}},\ }\href
  {https://doi.org/10.1063/5.0067660} {\bibfield  {journal} {\bibinfo
  {journal} {The Journal of Chemical Physics}\ }\textbf {\bibinfo {volume}
  {155}},\ \bibinfo {pages} {174115} (\bibinfo {year} {2021})}\BibitemShut
  {NoStop}%
\bibitem [{\citenamefont {Head-Marsden}\ \emph {et~al.}(2021)\citenamefont
  {Head-Marsden}, \citenamefont {Flick}, \citenamefont {Ciccarino},\ and\
  \citenamefont {Narang}}]{Head-Marsden.2021}%
  \BibitemOpen
  \bibfield  {author} {\bibinfo {author} {\bibfnamefont {K.}~\bibnamefont
  {Head-Marsden}}, \bibinfo {author} {\bibfnamefont {J.}~\bibnamefont {Flick}},
  \bibinfo {author} {\bibfnamefont {C.~J.}\ \bibnamefont {Ciccarino}},\ and\
  \bibinfo {author} {\bibfnamefont {P.}~\bibnamefont {Narang}},\ }\bibfield
  {title} {\bibinfo {title} {{Quantum Information and Algorithms for Correlated
  Quantum Matter}},\ }\href {https://doi.org/10.1021/acs.chemrev.0c00620}
  {\bibfield  {journal} {\bibinfo  {journal} {Chemical Reviews}\ }\textbf
  {\bibinfo {volume} {121}},\ \bibinfo {pages} {3061} (\bibinfo {year}
  {2021})}\BibitemShut {NoStop}%
\bibitem [{\citenamefont {Bharti}\ \emph {et~al.}(2022)\citenamefont {Bharti},
  \citenamefont {Cervera-Lierta}, \citenamefont {Kyaw}, \citenamefont {Haug},
  \citenamefont {Alperin-Lea}, \citenamefont {Anand}, \citenamefont {Degroote},
  \citenamefont {Heimonen}, \citenamefont {Kottmann}, \citenamefont {Menke},
  \citenamefont {Mok}, \citenamefont {Sim}, \citenamefont {Kwek},\ and\
  \citenamefont {Aspuru-Guzik}}]{Bharti.2022}%
  \BibitemOpen
  \bibfield  {author} {\bibinfo {author} {\bibfnamefont {K.}~\bibnamefont
  {Bharti}}, \bibinfo {author} {\bibfnamefont {A.}~\bibnamefont
  {Cervera-Lierta}}, \bibinfo {author} {\bibfnamefont {T.~H.}\ \bibnamefont
  {Kyaw}}, \bibinfo {author} {\bibfnamefont {T.}~\bibnamefont {Haug}}, \bibinfo
  {author} {\bibfnamefont {S.}~\bibnamefont {Alperin-Lea}}, \bibinfo {author}
  {\bibfnamefont {A.}~\bibnamefont {Anand}}, \bibinfo {author} {\bibfnamefont
  {M.}~\bibnamefont {Degroote}}, \bibinfo {author} {\bibfnamefont
  {H.}~\bibnamefont {Heimonen}}, \bibinfo {author} {\bibfnamefont {J.~S.}\
  \bibnamefont {Kottmann}}, \bibinfo {author} {\bibfnamefont {T.}~\bibnamefont
  {Menke}}, \bibinfo {author} {\bibfnamefont {W.-K.}\ \bibnamefont {Mok}},
  \bibinfo {author} {\bibfnamefont {S.}~\bibnamefont {Sim}}, \bibinfo {author}
  {\bibfnamefont {L.-C.}\ \bibnamefont {Kwek}},\ and\ \bibinfo {author}
  {\bibfnamefont {A.}~\bibnamefont {Aspuru-Guzik}},\ }\bibfield  {title}
  {\bibinfo {title} {{Noisy intermediate-scale quantum algorithms}},\ }\href
  {https://doi.org/10.1103/revmodphys.94.015004} {\bibfield  {journal}
  {\bibinfo  {journal} {Reviews of Modern Physics}\ }\textbf {\bibinfo {volume}
  {94}},\ \bibinfo {pages} {015004} (\bibinfo {year} {2022})}\BibitemShut
  {NoStop}%
\bibitem [{\citenamefont {Lloyd}(1993)}]{Lloyd.1993}%
  \BibitemOpen
  \bibfield  {author} {\bibinfo {author} {\bibfnamefont {S.}~\bibnamefont
  {Lloyd}},\ }\bibfield  {title} {\bibinfo {title} {{A potentially realizable
  quantum computer.}},\ }\href {https://doi.org/10.1126/science.261.5128.1569}
  {\bibfield  {journal} {\bibinfo  {journal} {Science (New York, N.Y.)}\
  }\textbf {\bibinfo {volume} {261}},\ \bibinfo {pages} {1569} (\bibinfo {year}
  {1993})}\BibitemShut {NoStop}%
\bibitem [{\citenamefont {Gao}\ \emph {et~al.}(2021)\citenamefont {Gao},
  \citenamefont {Jones}, \citenamefont {Motta}, \citenamefont {Sugawara},
  \citenamefont {Watanabe}, \citenamefont {Kobayashi}, \citenamefont
  {Watanabe}, \citenamefont {Ohnishi}, \citenamefont {Nakamura},\ and\
  \citenamefont {Yamamoto}}]{Gao.2021}%
  \BibitemOpen
  \bibfield  {author} {\bibinfo {author} {\bibfnamefont {Q.}~\bibnamefont
  {Gao}}, \bibinfo {author} {\bibfnamefont {G.~O.}\ \bibnamefont {Jones}},
  \bibinfo {author} {\bibfnamefont {M.}~\bibnamefont {Motta}}, \bibinfo
  {author} {\bibfnamefont {M.}~\bibnamefont {Sugawara}}, \bibinfo {author}
  {\bibfnamefont {H.~C.}\ \bibnamefont {Watanabe}}, \bibinfo {author}
  {\bibfnamefont {T.}~\bibnamefont {Kobayashi}}, \bibinfo {author}
  {\bibfnamefont {E.}~\bibnamefont {Watanabe}}, \bibinfo {author}
  {\bibfnamefont {Y.-y.}\ \bibnamefont {Ohnishi}}, \bibinfo {author}
  {\bibfnamefont {H.}~\bibnamefont {Nakamura}},\ and\ \bibinfo {author}
  {\bibfnamefont {N.}~\bibnamefont {Yamamoto}},\ }\bibfield  {title} {\bibinfo
  {title} {{Applications of quantum computing for investigations of electronic
  transitions in phenylsulfonyl-carbazole TADF emitters}},\ }\href
  {https://doi.org/10.1038/s41524-021-00540-6} {\bibfield  {journal} {\bibinfo
  {journal} {npj Computational Materials}\ }\textbf {\bibinfo {volume} {7}},\
  \bibinfo {pages} {70} (\bibinfo {year} {2021})},\ \Eprint
  {https://arxiv.org/abs/2007.15795} {2007.15795} \BibitemShut {NoStop}%
\bibitem [{\citenamefont {Asthana}\ \emph {et~al.}(2022)\citenamefont
  {Asthana}, \citenamefont {Kumar}, \citenamefont {Abraham}, \citenamefont
  {Grimsley}, \citenamefont {Zhang}, \citenamefont {Cincio}, \citenamefont
  {Tretiak}, \citenamefont {Dub}, \citenamefont {Economou}, \citenamefont
  {Barnes},\ and\ \citenamefont {Mayhall}}]{Asthana.2022}%
  \BibitemOpen
  \bibfield  {author} {\bibinfo {author} {\bibfnamefont {A.}~\bibnamefont
  {Asthana}}, \bibinfo {author} {\bibfnamefont {A.}~\bibnamefont {Kumar}},
  \bibinfo {author} {\bibfnamefont {V.}~\bibnamefont {Abraham}}, \bibinfo
  {author} {\bibfnamefont {H.}~\bibnamefont {Grimsley}}, \bibinfo {author}
  {\bibfnamefont {Y.}~\bibnamefont {Zhang}}, \bibinfo {author} {\bibfnamefont
  {L.}~\bibnamefont {Cincio}}, \bibinfo {author} {\bibfnamefont
  {S.}~\bibnamefont {Tretiak}}, \bibinfo {author} {\bibfnamefont {P.~A.}\
  \bibnamefont {Dub}}, \bibinfo {author} {\bibfnamefont {S.~E.}\ \bibnamefont
  {Economou}}, \bibinfo {author} {\bibfnamefont {E.}~\bibnamefont {Barnes}},\
  and\ \bibinfo {author} {\bibfnamefont {N.~J.}\ \bibnamefont {Mayhall}},\
  }\bibfield  {title} {\bibinfo {title} {{Equation-of-motion variational
  quantum eigensolver method for computing molecular excitation energies,
  ionization potentials, and electron affinities}},\ }\bibfield  {journal}
  {\bibinfo  {journal} {arXiv}\ }\href
  {https://doi.org/10.48550/arxiv.2206.10502} {10.48550/arxiv.2206.10502}
  (\bibinfo {year} {2022}),\ \Eprint {https://arxiv.org/abs/2206.10502}
  {2206.10502} \BibitemShut {NoStop}%
\bibitem [{\citenamefont {Kumar}\ \emph {et~al.}(2022)\citenamefont {Kumar},
  \citenamefont {Asthana}, \citenamefont {Masteran}, \citenamefont {Valeev},
  \citenamefont {Zhang}, \citenamefont {Cincio}, \citenamefont {Tretiak},\ and\
  \citenamefont {Dub}}]{Kumar.2022}%
  \BibitemOpen
  \bibfield  {author} {\bibinfo {author} {\bibfnamefont {A.}~\bibnamefont
  {Kumar}}, \bibinfo {author} {\bibfnamefont {A.}~\bibnamefont {Asthana}},
  \bibinfo {author} {\bibfnamefont {C.}~\bibnamefont {Masteran}}, \bibinfo
  {author} {\bibfnamefont {E.~F.}\ \bibnamefont {Valeev}}, \bibinfo {author}
  {\bibfnamefont {Y.}~\bibnamefont {Zhang}}, \bibinfo {author} {\bibfnamefont
  {L.}~\bibnamefont {Cincio}}, \bibinfo {author} {\bibfnamefont
  {S.}~\bibnamefont {Tretiak}},\ and\ \bibinfo {author} {\bibfnamefont {P.~A.}\
  \bibnamefont {Dub}},\ }\bibfield  {title} {\bibinfo {title} {{Quantum
  Simulation of Molecular Electronic States with a Transcorrelated Hamiltonian:
  Higher Accuracy with Fewer Qubits}},\ }\href
  {https://doi.org/10.1021/acs.jctc.2c00520} {\bibfield  {journal} {\bibinfo
  {journal} {Journal of Chemical Theory and Computation}\ }\textbf {\bibinfo
  {volume} {18}},\ \bibinfo {pages} {5312} (\bibinfo {year}
  {2022})}\BibitemShut {NoStop}%
\bibitem [{\citenamefont {Hlatshwayo}\ \emph {et~al.}(2022)\citenamefont
  {Hlatshwayo}, \citenamefont {Zhang}, \citenamefont {Wibowo}, \citenamefont
  {LaRose}, \citenamefont {Lacroix},\ and\ \citenamefont
  {Litvinova}}]{Hlatshwayo.2022}%
  \BibitemOpen
  \bibfield  {author} {\bibinfo {author} {\bibfnamefont {M.~Q.}\ \bibnamefont
  {Hlatshwayo}}, \bibinfo {author} {\bibfnamefont {Y.}~\bibnamefont {Zhang}},
  \bibinfo {author} {\bibfnamefont {H.}~\bibnamefont {Wibowo}}, \bibinfo
  {author} {\bibfnamefont {R.}~\bibnamefont {LaRose}}, \bibinfo {author}
  {\bibfnamefont {D.}~\bibnamefont {Lacroix}},\ and\ \bibinfo {author}
  {\bibfnamefont {E.}~\bibnamefont {Litvinova}},\ }\bibfield  {title} {\bibinfo
  {title} {{Simulating excited states of the Lipkin model on a quantum
  computer}},\ }\href {https://doi.org/10.1103/physrevc.106.024319} {\bibfield
  {journal} {\bibinfo  {journal} {Physical Review C}\ }\textbf {\bibinfo
  {volume} {106}},\ \bibinfo {pages} {024319} (\bibinfo {year} {2022})},\
  \Eprint {https://arxiv.org/abs/2203.01478} {2203.01478} \BibitemShut
  {NoStop}%
\bibitem [{\citenamefont {Asthana}\ \emph {et~al.}(2023)\citenamefont
  {Asthana}, \citenamefont {Kumar}, \citenamefont {Abraham}, \citenamefont
  {Grimsley}, \citenamefont {Zhang}, \citenamefont {Cincio}, \citenamefont
  {Tretiak}, \citenamefont {Dub}, \citenamefont {Economou}, \citenamefont
  {Barnes},\ and\ \citenamefont {Mayhall}}]{Asthana.2023}%
  \BibitemOpen
  \bibfield  {author} {\bibinfo {author} {\bibfnamefont {A.}~\bibnamefont
  {Asthana}}, \bibinfo {author} {\bibfnamefont {A.}~\bibnamefont {Kumar}},
  \bibinfo {author} {\bibfnamefont {V.}~\bibnamefont {Abraham}}, \bibinfo
  {author} {\bibfnamefont {H.}~\bibnamefont {Grimsley}}, \bibinfo {author}
  {\bibfnamefont {Y.}~\bibnamefont {Zhang}}, \bibinfo {author} {\bibfnamefont
  {L.}~\bibnamefont {Cincio}}, \bibinfo {author} {\bibfnamefont
  {S.}~\bibnamefont {Tretiak}}, \bibinfo {author} {\bibfnamefont {P.~A.}\
  \bibnamefont {Dub}}, \bibinfo {author} {\bibfnamefont {S.~E.}\ \bibnamefont
  {Economou}}, \bibinfo {author} {\bibfnamefont {E.}~\bibnamefont {Barnes}},\
  and\ \bibinfo {author} {\bibfnamefont {N.~J.}\ \bibnamefont {Mayhall}},\
  }\bibfield  {title} {\bibinfo {title} {{Quantum self-consistent
  equation-of-motion method for computing molecular excitation energies,
  ionization potentials, and electron affinities on a quantum computer}},\
  }\href {https://doi.org/10.1039/d2sc05371c} {\bibfield  {journal} {\bibinfo
  {journal} {Chemical Science}\ }\textbf {\bibinfo {volume} {14}},\ \bibinfo
  {pages} {2405} (\bibinfo {year} {2023})}\BibitemShut {NoStop}%
\bibitem [{\citenamefont {Kim}\ and\ \citenamefont {Krylov}(2023)}]{Kim.2023}%
  \BibitemOpen
  \bibfield  {author} {\bibinfo {author} {\bibfnamefont {Y.}~\bibnamefont
  {Kim}}\ and\ \bibinfo {author} {\bibfnamefont {A.}~\bibnamefont {Krylov}},\
  }\bibfield  {title} {\bibinfo {title} {{Two algorithms for excited-states
  quantum solvers: Theory and application to EOM-UCCSD}},\ }\bibfield
  {journal} {\bibinfo  {journal} {ChemRxiv}\ }\href
  {https://doi.org/10.26434/chemrxiv-2023-fml2k} {10.26434/chemrxiv-2023-fml2k}
  (\bibinfo {year} {2023})\BibitemShut {NoStop}%
\bibitem [{\citenamefont {McClean}\ \emph {et~al.}(2017)\citenamefont
  {McClean}, \citenamefont {Kimchi-Schwartz}, \citenamefont {Carter},\ and\
  \citenamefont {Jong}}]{McClean.2017}%
  \BibitemOpen
  \bibfield  {author} {\bibinfo {author} {\bibfnamefont {J.~R.}\ \bibnamefont
  {McClean}}, \bibinfo {author} {\bibfnamefont {M.~E.}\ \bibnamefont
  {Kimchi-Schwartz}}, \bibinfo {author} {\bibfnamefont {J.}~\bibnamefont
  {Carter}},\ and\ \bibinfo {author} {\bibfnamefont {W.~A.~d.}\ \bibnamefont
  {Jong}},\ }\bibfield  {title} {\bibinfo {title} {{Hybrid quantum-classical
  hierarchy for mitigation of decoherence and determination of excited
  states}},\ }\href {https://doi.org/10.1103/physreva.95.042308} {\bibfield
  {journal} {\bibinfo  {journal} {Physical Review A}\ }\textbf {\bibinfo
  {volume} {95}},\ \bibinfo {pages} {042308} (\bibinfo {year} {2017})},\
  \Eprint {https://arxiv.org/abs/1603.05681} {1603.05681} \BibitemShut
  {NoStop}%
\bibitem [{\citenamefont {Colless}\ \emph {et~al.}(2018)\citenamefont
  {Colless}, \citenamefont {Ramasesh}, \citenamefont {Dahlen}, \citenamefont
  {Blok}, \citenamefont {Kimchi-Schwartz}, \citenamefont {McClean},
  \citenamefont {Carter}, \citenamefont {Jong},\ and\ \citenamefont
  {Siddiqi}}]{Colless.2018}%
  \BibitemOpen
  \bibfield  {author} {\bibinfo {author} {\bibfnamefont {J.~I.}\ \bibnamefont
  {Colless}}, \bibinfo {author} {\bibfnamefont {V.~V.}\ \bibnamefont
  {Ramasesh}}, \bibinfo {author} {\bibfnamefont {D.}~\bibnamefont {Dahlen}},
  \bibinfo {author} {\bibfnamefont {M.~S.}\ \bibnamefont {Blok}}, \bibinfo
  {author} {\bibfnamefont {M.~E.}\ \bibnamefont {Kimchi-Schwartz}}, \bibinfo
  {author} {\bibfnamefont {J.~R.}\ \bibnamefont {McClean}}, \bibinfo {author}
  {\bibfnamefont {J.}~\bibnamefont {Carter}}, \bibinfo {author} {\bibfnamefont
  {W.~A.~d.}\ \bibnamefont {Jong}},\ and\ \bibinfo {author} {\bibfnamefont
  {I.}~\bibnamefont {Siddiqi}},\ }\bibfield  {title} {\bibinfo {title}
  {{Computation of Molecular Spectra on a Quantum Processor with an
  Error-Resilient Algorithm}},\ }\href
  {https://doi.org/10.1103/physrevx.8.011021} {\bibfield  {journal} {\bibinfo
  {journal} {Physical Review X}\ }\textbf {\bibinfo {volume} {8}},\ \bibinfo
  {pages} {011021} (\bibinfo {year} {2018})},\ \Eprint
  {https://arxiv.org/abs/1707.06408} {1707.06408} \BibitemShut {NoStop}%
\bibitem [{\citenamefont {Nakanishi}\ \emph {et~al.}(2019)\citenamefont
  {Nakanishi}, \citenamefont {Mitarai},\ and\ \citenamefont
  {Fujii}}]{Nakanishi.2019}%
  \BibitemOpen
  \bibfield  {author} {\bibinfo {author} {\bibfnamefont {K.~M.}\ \bibnamefont
  {Nakanishi}}, \bibinfo {author} {\bibfnamefont {K.}~\bibnamefont {Mitarai}},\
  and\ \bibinfo {author} {\bibfnamefont {K.}~\bibnamefont {Fujii}},\ }\bibfield
   {title} {\bibinfo {title} {{Subspace-search variational quantum eigensolver
  for excited states}},\ }\href
  {https://doi.org/10.1103/physrevresearch.1.033062} {\bibfield  {journal}
  {\bibinfo  {journal} {Physical Review Research}\ }\textbf {\bibinfo {volume}
  {1}},\ \bibinfo {pages} {033062} (\bibinfo {year} {2019})},\ \Eprint
  {https://arxiv.org/abs/1810.09434} {1810.09434} \BibitemShut {NoStop}%
\bibitem [{\citenamefont {Bian}\ \emph {et~al.}(2019)\citenamefont {Bian},
  \citenamefont {Murphy}, \citenamefont {Xia}, \citenamefont {Daskin},\ and\
  \citenamefont {Kais}}]{Bian.2019}%
  \BibitemOpen
  \bibfield  {author} {\bibinfo {author} {\bibfnamefont {T.}~\bibnamefont
  {Bian}}, \bibinfo {author} {\bibfnamefont {D.}~\bibnamefont {Murphy}},
  \bibinfo {author} {\bibfnamefont {R.}~\bibnamefont {Xia}}, \bibinfo {author}
  {\bibfnamefont {A.}~\bibnamefont {Daskin}},\ and\ \bibinfo {author}
  {\bibfnamefont {S.}~\bibnamefont {Kais}},\ }\bibfield  {title} {\bibinfo
  {title} {{Quantum computing methods for electronic states of the water
  molecule}},\ }\href {https://doi.org/10.1080/00268976.2019.1580392}
  {\bibfield  {journal} {\bibinfo  {journal} {Molecular Physics}\ }\textbf
  {\bibinfo {volume} {117}},\ \bibinfo {pages} {2069} (\bibinfo {year}
  {2019})}\BibitemShut {NoStop}%
\bibitem [{\citenamefont {Motta}\ \emph {et~al.}(2020)\citenamefont {Motta},
  \citenamefont {Sun}, \citenamefont {Tan}, \citenamefont {O’Rourke},
  \citenamefont {Ye}, \citenamefont {Minnich}, \citenamefont {Brandão},\ and\
  \citenamefont {Chan}}]{Motta.2020}%
  \BibitemOpen
  \bibfield  {author} {\bibinfo {author} {\bibfnamefont {M.}~\bibnamefont
  {Motta}}, \bibinfo {author} {\bibfnamefont {C.}~\bibnamefont {Sun}}, \bibinfo
  {author} {\bibfnamefont {A.~T.~K.}\ \bibnamefont {Tan}}, \bibinfo {author}
  {\bibfnamefont {M.~J.}\ \bibnamefont {O’Rourke}}, \bibinfo {author}
  {\bibfnamefont {E.}~\bibnamefont {Ye}}, \bibinfo {author} {\bibfnamefont
  {A.~J.}\ \bibnamefont {Minnich}}, \bibinfo {author} {\bibfnamefont {F.~G.
  S.~L.}\ \bibnamefont {Brandão}},\ and\ \bibinfo {author} {\bibfnamefont
  {G.~K.-L.}\ \bibnamefont {Chan}},\ }\bibfield  {title} {\bibinfo {title}
  {{Determining eigenstates and thermal states on a quantum computer using
  quantum imaginary time evolution}},\ }\href
  {https://doi.org/10.1038/s41567-019-0704-4} {\bibfield  {journal} {\bibinfo
  {journal} {Nature Physics}\ }\textbf {\bibinfo {volume} {16}},\ \bibinfo
  {pages} {205} (\bibinfo {year} {2020})},\ \Eprint
  {https://arxiv.org/abs/1901.07653} {1901.07653} \BibitemShut {NoStop}%
\bibitem [{\citenamefont {McClean}\ \emph {et~al.}(2020)\citenamefont
  {McClean}, \citenamefont {Jiang}, \citenamefont {Rubin}, \citenamefont
  {Babbush},\ and\ \citenamefont {Neven}}]{McClean.2020}%
  \BibitemOpen
  \bibfield  {author} {\bibinfo {author} {\bibfnamefont {J.~R.}\ \bibnamefont
  {McClean}}, \bibinfo {author} {\bibfnamefont {Z.}~\bibnamefont {Jiang}},
  \bibinfo {author} {\bibfnamefont {N.~C.}\ \bibnamefont {Rubin}}, \bibinfo
  {author} {\bibfnamefont {R.}~\bibnamefont {Babbush}},\ and\ \bibinfo {author}
  {\bibfnamefont {H.}~\bibnamefont {Neven}},\ }\bibfield  {title} {\bibinfo
  {title} {{Decoding quantum errors with subspace expansions}},\ }\href
  {https://doi.org/10.1038/s41467-020-14341-w} {\bibfield  {journal} {\bibinfo
  {journal} {Nature Communications}\ }\textbf {\bibinfo {volume} {11}},\
  \bibinfo {pages} {636} (\bibinfo {year} {2020})},\ \Eprint
  {https://arxiv.org/abs/1903.05786} {1903.05786} \BibitemShut {NoStop}%
\bibitem [{\citenamefont {Takeshita}\ \emph {et~al.}(2020)\citenamefont
  {Takeshita}, \citenamefont {Rubin}, \citenamefont {Jiang}, \citenamefont
  {Lee}, \citenamefont {Babbush},\ and\ \citenamefont
  {McClean}}]{Takeshita.2020}%
  \BibitemOpen
  \bibfield  {author} {\bibinfo {author} {\bibfnamefont {T.}~\bibnamefont
  {Takeshita}}, \bibinfo {author} {\bibfnamefont {N.~C.}\ \bibnamefont
  {Rubin}}, \bibinfo {author} {\bibfnamefont {Z.}~\bibnamefont {Jiang}},
  \bibinfo {author} {\bibfnamefont {E.}~\bibnamefont {Lee}}, \bibinfo {author}
  {\bibfnamefont {R.}~\bibnamefont {Babbush}},\ and\ \bibinfo {author}
  {\bibfnamefont {J.~R.}\ \bibnamefont {McClean}},\ }\bibfield  {title}
  {\bibinfo {title} {{Increasing the Representation Accuracy of Quantum
  Simulations of Chemistry without Extra Quantum Resources}},\ }\href
  {https://doi.org/10.1103/physrevx.10.011004} {\bibfield  {journal} {\bibinfo
  {journal} {Physical Review X}\ }\textbf {\bibinfo {volume} {10}},\ \bibinfo
  {pages} {011004} (\bibinfo {year} {2020})},\ \Eprint
  {https://arxiv.org/abs/1902.10679} {1902.10679} \BibitemShut {NoStop}%
\bibitem [{\citenamefont {Francis}\ \emph {et~al.}(2022)\citenamefont
  {Francis}, \citenamefont {Agrawal}, \citenamefont {Howard}, \citenamefont
  {Kökcü},\ and\ \citenamefont {Kemper}}]{Francis.2022}%
  \BibitemOpen
  \bibfield  {author} {\bibinfo {author} {\bibfnamefont {A.}~\bibnamefont
  {Francis}}, \bibinfo {author} {\bibfnamefont {A.~A.}\ \bibnamefont
  {Agrawal}}, \bibinfo {author} {\bibfnamefont {J.~H.}\ \bibnamefont {Howard}},
  \bibinfo {author} {\bibfnamefont {E.}~\bibnamefont {Kökcü}},\ and\ \bibinfo
  {author} {\bibfnamefont {A.~F.}\ \bibnamefont {Kemper}},\ }\bibfield  {title}
  {\bibinfo {title} {{Subspace Diagonalization on Quantum Computers using
  Eigenvector Continuation}},\ }\bibfield  {journal} {\bibinfo  {journal}
  {arXiv}\ }\href {https://doi.org/10.48550/arxiv.2209.10571}
  {10.48550/arxiv.2209.10571} (\bibinfo {year} {2022}),\ \Eprint
  {https://arxiv.org/abs/2209.10571} {2209.10571} \BibitemShut {NoStop}%
\bibitem [{\citenamefont {Huang}\ \emph {et~al.}(2022)\citenamefont {Huang},
  \citenamefont {Govoni},\ and\ \citenamefont {Galli}}]{Huang.2022}%
  \BibitemOpen
  \bibfield  {author} {\bibinfo {author} {\bibfnamefont {B.}~\bibnamefont
  {Huang}}, \bibinfo {author} {\bibfnamefont {M.}~\bibnamefont {Govoni}},\ and\
  \bibinfo {author} {\bibfnamefont {G.}~\bibnamefont {Galli}},\ }\bibfield
  {title} {\bibinfo {title} {{Simulating the Electronic Structure of Spin
  Defects on Quantum Computers}},\ }\href
  {https://doi.org/10.1103/prxquantum.3.010339} {\bibfield  {journal} {\bibinfo
   {journal} {PRX Quantum}\ }\textbf {\bibinfo {volume} {3}},\ \bibinfo {pages}
  {010339} (\bibinfo {year} {2022})},\ \Eprint
  {https://arxiv.org/abs/2112.04435} {2112.04435} \BibitemShut {NoStop}%
\bibitem [{\citenamefont {Tkachenko}\ \emph {et~al.}(2022)\citenamefont
  {Tkachenko}, \citenamefont {Zhang}, \citenamefont {Cincio}, \citenamefont
  {Boldyrev}, \citenamefont {Tretiak},\ and\ \citenamefont
  {Dub}}]{Tkachenko.2022}%
  \BibitemOpen
  \bibfield  {author} {\bibinfo {author} {\bibfnamefont {N.~V.}\ \bibnamefont
  {Tkachenko}}, \bibinfo {author} {\bibfnamefont {Y.}~\bibnamefont {Zhang}},
  \bibinfo {author} {\bibfnamefont {L.}~\bibnamefont {Cincio}}, \bibinfo
  {author} {\bibfnamefont {A.~I.}\ \bibnamefont {Boldyrev}}, \bibinfo {author}
  {\bibfnamefont {S.}~\bibnamefont {Tretiak}},\ and\ \bibinfo {author}
  {\bibfnamefont {P.~A.}\ \bibnamefont {Dub}},\ }\bibfield  {title} {\bibinfo
  {title} {{Quantum Davidson Algorithm for Excited States}},\ }\bibfield
  {journal} {\bibinfo  {journal} {arXiv}\ }\href
  {https://doi.org/10.48550/arxiv.2204.10741} {10.48550/arxiv.2204.10741}
  (\bibinfo {year} {2022}),\ \Eprint {https://arxiv.org/abs/2204.10741}
  {2204.10741} \BibitemShut {NoStop}%
\bibitem [{\citenamefont {Cortes}\ and\ \citenamefont
  {Gray}(2022)}]{Cortes.2022}%
  \BibitemOpen
  \bibfield  {author} {\bibinfo {author} {\bibfnamefont {C.~L.}\ \bibnamefont
  {Cortes}}\ and\ \bibinfo {author} {\bibfnamefont {S.~K.}\ \bibnamefont
  {Gray}},\ }\bibfield  {title} {\bibinfo {title} {{Quantum Krylov subspace
  algorithms for ground- and excited-state energy estimation}},\ }\href
  {https://doi.org/10.1103/physreva.105.022417} {\bibfield  {journal} {\bibinfo
   {journal} {Physical Review A}\ }\textbf {\bibinfo {volume} {105}},\ \bibinfo
  {pages} {022417} (\bibinfo {year} {2022})},\ \Eprint
  {https://arxiv.org/abs/2109.06868} {2109.06868} \BibitemShut {NoStop}%
\bibitem [{\citenamefont {Shen}\ \emph {et~al.}(2022)\citenamefont {Shen},
  \citenamefont {Klymko}, \citenamefont {Sud}, \citenamefont {Williams-Young},
  \citenamefont {Jong},\ and\ \citenamefont {Tubman}}]{Shen.2022}%
  \BibitemOpen
  \bibfield  {author} {\bibinfo {author} {\bibfnamefont {Y.}~\bibnamefont
  {Shen}}, \bibinfo {author} {\bibfnamefont {K.}~\bibnamefont {Klymko}},
  \bibinfo {author} {\bibfnamefont {J.}~\bibnamefont {Sud}}, \bibinfo {author}
  {\bibfnamefont {D.~B.}\ \bibnamefont {Williams-Young}}, \bibinfo {author}
  {\bibfnamefont {W.~A.~d.}\ \bibnamefont {Jong}},\ and\ \bibinfo {author}
  {\bibfnamefont {N.~M.}\ \bibnamefont {Tubman}},\ }\bibfield  {title}
  {\bibinfo {title} {{Real-Time Krylov Theory for Quantum Computing
  Algorithms}},\ }\bibfield  {journal} {\bibinfo  {journal} {arXiv}\ }\href
  {https://doi.org/10.48550/arxiv.2208.01063} {10.48550/arxiv.2208.01063}
  (\bibinfo {year} {2022}),\ \Eprint {https://arxiv.org/abs/2208.01063}
  {2208.01063} \BibitemShut {NoStop}%
\bibitem [{\citenamefont {Yoshioka}\ \emph {et~al.}(2022)\citenamefont
  {Yoshioka}, \citenamefont {Hakoshima}, \citenamefont {Matsuzaki},
  \citenamefont {Tokunaga}, \citenamefont {Suzuki},\ and\ \citenamefont
  {Endo}}]{Yoshioka.2022}%
  \BibitemOpen
  \bibfield  {author} {\bibinfo {author} {\bibfnamefont {N.}~\bibnamefont
  {Yoshioka}}, \bibinfo {author} {\bibfnamefont {H.}~\bibnamefont {Hakoshima}},
  \bibinfo {author} {\bibfnamefont {Y.}~\bibnamefont {Matsuzaki}}, \bibinfo
  {author} {\bibfnamefont {Y.}~\bibnamefont {Tokunaga}}, \bibinfo {author}
  {\bibfnamefont {Y.}~\bibnamefont {Suzuki}},\ and\ \bibinfo {author}
  {\bibfnamefont {S.}~\bibnamefont {Endo}},\ }\bibfield  {title} {\bibinfo
  {title} {{Generalized Quantum Subspace Expansion}},\ }\href
  {https://doi.org/10.1103/physrevlett.129.020502} {\bibfield  {journal}
  {\bibinfo  {journal} {Physical Review Letters}\ }\textbf {\bibinfo {volume}
  {129}},\ \bibinfo {pages} {020502} (\bibinfo {year} {2022})},\ \Eprint
  {https://arxiv.org/abs/2107.02611} {2107.02611} \BibitemShut {NoStop}%
\bibitem [{\citenamefont {Motta}\ \emph {et~al.}(2023)\citenamefont {Motta},
  \citenamefont {Jones}, \citenamefont {Rice}, \citenamefont {Gujarati},
  \citenamefont {Sakuma}, \citenamefont {Liepuoniute}, \citenamefont {Garcia},\
  and\ \citenamefont {Ohnishi}}]{Motta.2023}%
  \BibitemOpen
  \bibfield  {author} {\bibinfo {author} {\bibfnamefont {M.}~\bibnamefont
  {Motta}}, \bibinfo {author} {\bibfnamefont {G.~O.}\ \bibnamefont {Jones}},
  \bibinfo {author} {\bibfnamefont {J.~E.}\ \bibnamefont {Rice}}, \bibinfo
  {author} {\bibfnamefont {T.~P.}\ \bibnamefont {Gujarati}}, \bibinfo {author}
  {\bibfnamefont {R.}~\bibnamefont {Sakuma}}, \bibinfo {author} {\bibfnamefont
  {I.}~\bibnamefont {Liepuoniute}}, \bibinfo {author} {\bibfnamefont {J.~M.}\
  \bibnamefont {Garcia}},\ and\ \bibinfo {author} {\bibfnamefont {Y.-y.}\
  \bibnamefont {Ohnishi}},\ }\bibfield  {title} {\bibinfo {title} {{Quantum
  chemistry simulation of ground- and excited-state properties of the sulfonium
  cation on a superconducting quantum processor}},\ }\href
  {https://doi.org/10.1039/d2sc06019a} {\bibfield  {journal} {\bibinfo
  {journal} {Chemical Science}\ }\textbf {\bibinfo {volume} {14}},\ \bibinfo
  {pages} {2915} (\bibinfo {year} {2023})},\ \Eprint
  {https://arxiv.org/abs/2208.02414} {2208.02414} \BibitemShut {NoStop}%
\bibitem [{\citenamefont {Kanno}\ \emph {et~al.}(2023)\citenamefont {Kanno},
  \citenamefont {Kohda}, \citenamefont {Imai}, \citenamefont {Koh},
  \citenamefont {Mitarai}, \citenamefont {Mizukami},\ and\ \citenamefont
  {Nakagawa}}]{Kanno.2023}%
  \BibitemOpen
  \bibfield  {author} {\bibinfo {author} {\bibfnamefont {K.}~\bibnamefont
  {Kanno}}, \bibinfo {author} {\bibfnamefont {M.}~\bibnamefont {Kohda}},
  \bibinfo {author} {\bibfnamefont {R.}~\bibnamefont {Imai}}, \bibinfo {author}
  {\bibfnamefont {S.}~\bibnamefont {Koh}}, \bibinfo {author} {\bibfnamefont
  {K.}~\bibnamefont {Mitarai}}, \bibinfo {author} {\bibfnamefont
  {W.}~\bibnamefont {Mizukami}},\ and\ \bibinfo {author} {\bibfnamefont
  {Y.~O.}\ \bibnamefont {Nakagawa}},\ }\bibfield  {title} {\bibinfo {title}
  {{Quantum-Selected Configuration Interaction: classical diagonalization of
  Hamiltonians in subspaces selected by quantum computers}},\ }\bibfield
  {journal} {\bibinfo  {journal} {arXiv}\ }\href
  {https://doi.org/10.48550/arxiv.2302.11320} {10.48550/arxiv.2302.11320}
  (\bibinfo {year} {2023}),\ \Eprint {https://arxiv.org/abs/2302.11320}
  {2302.11320} \BibitemShut {NoStop}%
\bibitem [{\citenamefont {Choi}\ and\ \citenamefont
  {Izmaylov}(2023)}]{Choi.2023}%
  \BibitemOpen
  \bibfield  {author} {\bibinfo {author} {\bibfnamefont {S.}~\bibnamefont
  {Choi}}\ and\ \bibinfo {author} {\bibfnamefont {A.~F.}\ \bibnamefont
  {Izmaylov}},\ }\bibfield  {title} {\bibinfo {title} {{Measurement
  optimization techniques for excited electronic states in near-term quantum
  computing algorithms}},\ }\bibfield  {journal} {\bibinfo  {journal} {arXiv}\
  }\href {https://doi.org/10.48550/arxiv.2302.11421}
  {10.48550/arxiv.2302.11421} (\bibinfo {year} {2023}),\ \Eprint
  {https://arxiv.org/abs/2302.11421} {2302.11421} \BibitemShut {NoStop}%
\bibitem [{\citenamefont {Mazziotti}(1998)}]{Mazziotti.1998e39}%
  \BibitemOpen
  \bibfield  {author} {\bibinfo {author} {\bibfnamefont {D.~A.}\ \bibnamefont
  {Mazziotti}},\ }\bibfield  {title} {\bibinfo {title} {{Contracted
  Schrödinger equation: Determining quantum energies and two-particle density
  matrices without wave functions}},\ }\href
  {https://doi.org/10.1103/physreva.57.4219} {\bibfield  {journal} {\bibinfo
  {journal} {Physical Review A}\ }\textbf {\bibinfo {volume} {57}},\ \bibinfo
  {pages} {4219} (\bibinfo {year} {1998})}\BibitemShut {NoStop}%
\bibitem [{\citenamefont {Nakatsuji}\ and\ \citenamefont
  {Yasuda}(1996)}]{Nakatsuji.1996}%
  \BibitemOpen
  \bibfield  {author} {\bibinfo {author} {\bibfnamefont {H.}~\bibnamefont
  {Nakatsuji}}\ and\ \bibinfo {author} {\bibfnamefont {K.}~\bibnamefont
  {Yasuda}},\ }\bibfield  {title} {\bibinfo {title} {{Direct Determination of
  the Quantum-Mechanical Density Matrix Using the Density Equation}},\ }\href
  {https://doi.org/10.1103/physrevlett.76.1039} {\bibfield  {journal} {\bibinfo
   {journal} {Physical Review Letters}\ }\textbf {\bibinfo {volume} {76}},\
  \bibinfo {pages} {1039} (\bibinfo {year} {1996})}\BibitemShut {NoStop}%
\bibitem [{\citenamefont {Colmenero}\ and\ \citenamefont
  {Valdemoro}(1993)}]{Colmenero.1993}%
  \BibitemOpen
  \bibfield  {author} {\bibinfo {author} {\bibfnamefont {F.}~\bibnamefont
  {Colmenero}}\ and\ \bibinfo {author} {\bibfnamefont {C.}~\bibnamefont
  {Valdemoro}},\ }\bibfield  {title} {\bibinfo {title} {{Approximating q-order
  reduced density matrices in terms of the lower-order ones. II.
  Applications}},\ }\href {https://doi.org/10.1103/physreva.47.979} {\bibfield
  {journal} {\bibinfo  {journal} {Physical Review A}\ }\textbf {\bibinfo
  {volume} {47}},\ \bibinfo {pages} {979} (\bibinfo {year} {1993})}\BibitemShut
  {NoStop}%
\bibitem [{\citenamefont {Mazziotti}(2006)}]{Mazziotti.20060v3}%
  \BibitemOpen
  \bibfield  {author} {\bibinfo {author} {\bibfnamefont {D.~A.}\ \bibnamefont
  {Mazziotti}},\ }\bibfield  {title} {\bibinfo {title} {{Anti-Hermitian
  Contracted Schrödinger Equation: Direct Determination of the Two-Electron
  Reduced Density Matrices of Many-Electron Molecules}},\ }\href
  {https://doi.org/10.1103/physrevlett.97.143002} {\bibfield  {journal}
  {\bibinfo  {journal} {Physical Review Letters}\ }\textbf {\bibinfo {volume}
  {97}},\ \bibinfo {pages} {143002} (\bibinfo {year} {2006})}\BibitemShut
  {NoStop}%
\bibitem [{\citenamefont {Boyn}\ and\ \citenamefont
  {Mazziotti}(2021)}]{Boyn.2021}%
  \BibitemOpen
  \bibfield  {author} {\bibinfo {author} {\bibfnamefont {J.-N.}\ \bibnamefont
  {Boyn}}\ and\ \bibinfo {author} {\bibfnamefont {D.~A.}\ \bibnamefont
  {Mazziotti}},\ }\bibfield  {title} {\bibinfo {title} {{Accurate
  singlet–triplet gaps in biradicals via the spin averaged anti-Hermitian
  contracted Schrödinger equation}},\ }\href
  {https://doi.org/10.1063/5.0045007} {\bibfield  {journal} {\bibinfo
  {journal} {The Journal of Chemical Physics}\ }\textbf {\bibinfo {volume}
  {154}},\ \bibinfo {pages} {134103} (\bibinfo {year} {2021})},\ \Eprint
  {https://arxiv.org/abs/2104.00626} {2104.00626} \BibitemShut {NoStop}%
\bibitem [{\citenamefont {Mazziotti}(2004)}]{Mazziotti.20049l}%
  \BibitemOpen
  \bibfield  {author} {\bibinfo {author} {\bibfnamefont {D.~A.}\ \bibnamefont
  {Mazziotti}},\ }\bibfield  {title} {\bibinfo {title} {{Exactness of wave
  functions from two-body exponential transformations in many-body quantum
  theory}},\ }\href {https://doi.org/10.1103/physreva.69.012507} {\bibfield
  {journal} {\bibinfo  {journal} {Physical Review A}\ }\textbf {\bibinfo
  {volume} {69}},\ \bibinfo {pages} {012507} (\bibinfo {year}
  {2004})}\BibitemShut {NoStop}%
\bibitem [{\citenamefont {Mazziotti}(2020)}]{Mazziotti.2020}%
  \BibitemOpen
  \bibfield  {author} {\bibinfo {author} {\bibfnamefont {D.~A.}\ \bibnamefont
  {Mazziotti}},\ }\bibfield  {title} {\bibinfo {title} {{Exact two-body
  expansion of the many-particle wave function}},\ }\href
  {https://doi.org/10.1103/physreva.102.030802} {\bibfield  {journal} {\bibinfo
   {journal} {Physical Review A}\ }\textbf {\bibinfo {volume} {102}},\ \bibinfo
  {pages} {030802} (\bibinfo {year} {2020})},\ \Eprint
  {https://arxiv.org/abs/2010.02191} {2010.02191} \BibitemShut {NoStop}%
\bibitem [{\citenamefont {Smart}\ and\ \citenamefont
  {Mazziotti}(2021)}]{Smart.2021}%
  \BibitemOpen
  \bibfield  {author} {\bibinfo {author} {\bibfnamefont {S.~E.}\ \bibnamefont
  {Smart}}\ and\ \bibinfo {author} {\bibfnamefont {D.~A.}\ \bibnamefont
  {Mazziotti}},\ }\bibfield  {title} {\bibinfo {title} {{Quantum Solver of
  Contracted Eigenvalue Equations for Scalable Molecular Simulations on Quantum
  Computing Devices}},\ }\href {https://doi.org/10.1103/physrevlett.126.070504}
  {\bibfield  {journal} {\bibinfo  {journal} {Physical Review Letters}\
  }\textbf {\bibinfo {volume} {126}},\ \bibinfo {pages} {070504} (\bibinfo
  {year} {2021})},\ \Eprint {https://arxiv.org/abs/2004.11416} {2004.11416}
  \BibitemShut {NoStop}%
\bibitem [{\citenamefont {Boyn}\ \emph {et~al.}(2021)\citenamefont {Boyn},
  \citenamefont {Lykhin}, \citenamefont {Smart}, \citenamefont {Gagliardi},\
  and\ \citenamefont {Mazziotti}}]{Boyn.2021u94}%
  \BibitemOpen
  \bibfield  {author} {\bibinfo {author} {\bibfnamefont {J.-N.}\ \bibnamefont
  {Boyn}}, \bibinfo {author} {\bibfnamefont {A.~O.}\ \bibnamefont {Lykhin}},
  \bibinfo {author} {\bibfnamefont {S.~E.}\ \bibnamefont {Smart}}, \bibinfo
  {author} {\bibfnamefont {L.}~\bibnamefont {Gagliardi}},\ and\ \bibinfo
  {author} {\bibfnamefont {D.~A.}\ \bibnamefont {Mazziotti}},\ }\bibfield
  {title} {\bibinfo {title} {{Quantum-classical hybrid algorithm for the
  simulation of all-electron correlation}},\ }\href
  {https://doi.org/10.1063/5.0074842} {\bibfield  {journal} {\bibinfo
  {journal} {The Journal of Chemical Physics}\ }\textbf {\bibinfo {volume}
  {155}},\ \bibinfo {pages} {244106} (\bibinfo {year} {2021})},\ \Eprint
  {https://arxiv.org/abs/2106.11972} {2106.11972} \BibitemShut {NoStop}%
\bibitem [{\citenamefont {Smart}\ and\ \citenamefont
  {Mazziotti}(2022{\natexlab{a}})}]{Smart.2022l2}%
  \BibitemOpen
  \bibfield  {author} {\bibinfo {author} {\bibfnamefont {S.~E.}\ \bibnamefont
  {Smart}}\ and\ \bibinfo {author} {\bibfnamefont {D.~A.}\ \bibnamefont
  {Mazziotti}},\ }\bibfield  {title} {\bibinfo {title} {{Accelerated
  Convergence of Contracted Quantum Eigensolvers through a Quasi-Second-Order,
  Locally Parameterized Optimization}},\ }\href
  {https://doi.org/10.1021/acs.jctc.2c00446} {\bibfield  {journal} {\bibinfo
  {journal} {Journal of Chemical Theory and Computation}\ }\textbf {\bibinfo
  {volume} {18}},\ \bibinfo {pages} {5286} (\bibinfo {year}
  {2022}{\natexlab{a}})}\BibitemShut {NoStop}%
\bibitem [{\citenamefont {Smart}\ \emph {et~al.}(2022)\citenamefont {Smart},
  \citenamefont {Boyn},\ and\ \citenamefont {Mazziotti}}]{Smart.2022w8u}%
  \BibitemOpen
  \bibfield  {author} {\bibinfo {author} {\bibfnamefont {S.~E.}\ \bibnamefont
  {Smart}}, \bibinfo {author} {\bibfnamefont {J.-N.}\ \bibnamefont {Boyn}},\
  and\ \bibinfo {author} {\bibfnamefont {D.~A.}\ \bibnamefont {Mazziotti}},\
  }\bibfield  {title} {\bibinfo {title} {{Resolving correlated states of
  benzyne with an error-mitigated contracted quantum eigensolver}},\ }\href
  {https://doi.org/10.1103/physreva.105.022405} {\bibfield  {journal} {\bibinfo
   {journal} {Physical Review A}\ }\textbf {\bibinfo {volume} {105}},\ \bibinfo
  {pages} {022405} (\bibinfo {year} {2022})},\ \Eprint
  {https://arxiv.org/abs/2103.06876} {2103.06876} \BibitemShut {NoStop}%
\bibitem [{\citenamefont {Smart}\ and\ \citenamefont
  {Mazziotti}(2022{\natexlab{b}})}]{Smart.2022}%
  \BibitemOpen
  \bibfield  {author} {\bibinfo {author} {\bibfnamefont {S.~E.}\ \bibnamefont
  {Smart}}\ and\ \bibinfo {author} {\bibfnamefont {D.~A.}\ \bibnamefont
  {Mazziotti}},\ }\bibfield  {title} {\bibinfo {title} {{Many-fermion
  simulation from the contracted quantum eigensolver without fermionic encoding
  of the wave function}},\ }\href {https://doi.org/10.1103/physreva.105.062424}
  {\bibfield  {journal} {\bibinfo  {journal} {Physical Review A}\ }\textbf
  {\bibinfo {volume} {105}},\ \bibinfo {pages} {062424} (\bibinfo {year}
  {2022}{\natexlab{b}})},\ \Eprint {https://arxiv.org/abs/2205.01725}
  {2205.01725} \BibitemShut {NoStop}%
\bibitem [{\citenamefont {Smart}\ and\ \citenamefont
  {Mazziotti}(2023)}]{Smart.2023}%
  \BibitemOpen
  \bibfield  {author} {\bibinfo {author} {\bibfnamefont {S.~E.}\ \bibnamefont
  {Smart}}\ and\ \bibinfo {author} {\bibfnamefont {D.~A.}\ \bibnamefont
  {Mazziotti}},\ }\bibfield  {title} {\bibinfo {title} {{Verifiably Exact
  Solution of the Electronic Schrodinger Equation on Quantum Devices}},\
  }\bibfield  {journal} {\bibinfo  {journal} {arXiv}\ }\href
  {https://doi.org/10.48550/arxiv.2303.00758} {10.48550/arxiv.2303.00758}
  (\bibinfo {year} {2023}),\ \Eprint {https://arxiv.org/abs/2303.00758}
  {2303.00758} \BibitemShut {NoStop}%
\bibitem [{\citenamefont {Nakatsuji}(1976)}]{Nakatsuji.1976}%
  \BibitemOpen
  \bibfield  {author} {\bibinfo {author} {\bibfnamefont {H.}~\bibnamefont
  {Nakatsuji}},\ }\bibfield  {title} {\bibinfo {title} {{Equation for the
  direct determination of the density matrix}},\ }\href
  {https://doi.org/10.1103/physreva.14.41} {\bibfield  {journal} {\bibinfo
  {journal} {Physical Review A}\ }\textbf {\bibinfo {volume} {14}},\ \bibinfo
  {pages} {41} (\bibinfo {year} {1976})}\BibitemShut {NoStop}%
\bibitem [{\citenamefont {Mazziotti}(2007{\natexlab{a}})}]{Mazziotti.2007k2h}%
  \BibitemOpen
  \bibfield  {author} {\bibinfo {author} {\bibfnamefont {D.~A.}\ \bibnamefont
  {Mazziotti}},\ }\bibfield  {title} {\bibinfo {title} {{Multireference
  many-electron correlation energies from two-electron reduced density matrices
  computed by solving the anti-Hermitian contracted Schrödinger equation}},\
  }\href {https://doi.org/10.1103/physreva.76.052502} {\bibfield  {journal}
  {\bibinfo  {journal} {Physical Review A}\ }\textbf {\bibinfo {volume} {76}},\
  \bibinfo {pages} {052502} (\bibinfo {year} {2007}{\natexlab{a}})}\BibitemShut
  {NoStop}%
\bibitem [{\citenamefont {Mazziotti}(2007{\natexlab{b}})}]{Mazziotti.2007}%
  \BibitemOpen
  \bibfield  {author} {\bibinfo {author} {\bibfnamefont {D.~A.}\ \bibnamefont
  {Mazziotti}},\ }\bibfield  {title} {\bibinfo {title} {{Anti-Hermitian part of
  the contracted Schrödinger equation for the direct calculation of
  two-electron reduced density matrices}},\ }\href
  {https://doi.org/10.1103/physreva.75.022505} {\bibfield  {journal} {\bibinfo
  {journal} {Physical Review A}\ }\textbf {\bibinfo {volume} {75}},\ \bibinfo
  {pages} {022505} (\bibinfo {year} {2007}{\natexlab{b}})}\BibitemShut
  {NoStop}%
\bibitem [{\citenamefont {Gidofalvi}\ and\ \citenamefont
  {Mazziotti}(2009)}]{Gidofalvi.2009}%
  \BibitemOpen
  \bibfield  {author} {\bibinfo {author} {\bibfnamefont {G.}~\bibnamefont
  {Gidofalvi}}\ and\ \bibinfo {author} {\bibfnamefont {D.~A.}\ \bibnamefont
  {Mazziotti}},\ }\bibfield  {title} {\bibinfo {title} {{Direct calculation of
  excited-state electronic energies and two-electron reduced density matrices
  from the anti-Hermitian contracted Schrödinger equation}},\ }\href
  {https://doi.org/10.1103/physreva.80.022507} {\bibfield  {journal} {\bibinfo
  {journal} {Physical Review A}\ }\textbf {\bibinfo {volume} {80}},\ \bibinfo
  {pages} {022507} (\bibinfo {year} {2009})}\BibitemShut {NoStop}%
\bibitem [{\citenamefont {Snyder}\ \emph {et~al.}(2010)\citenamefont {Snyder},
  \citenamefont {Rothman}, \citenamefont {Foley},\ and\ \citenamefont
  {Mazziotti}}]{Snyder.2010}%
  \BibitemOpen
  \bibfield  {author} {\bibinfo {author} {\bibfnamefont {J.~W.}\ \bibnamefont
  {Snyder}}, \bibinfo {author} {\bibfnamefont {A.~E.}\ \bibnamefont {Rothman}},
  \bibinfo {author} {\bibfnamefont {J.~J.}\ \bibnamefont {Foley}},\ and\
  \bibinfo {author} {\bibfnamefont {D.~A.}\ \bibnamefont {Mazziotti}},\
  }\bibfield  {title} {\bibinfo {title} {{Conical intersections in triplet
  excited states of methylene from the anti-Hermitian contracted Schrödinger
  equation}},\ }\href {https://doi.org/10.1063/1.3394020} {\bibfield  {journal}
  {\bibinfo  {journal} {The Journal of Chemical Physics}\ }\textbf {\bibinfo
  {volume} {132}},\ \bibinfo {pages} {154109} (\bibinfo {year}
  {2010})}\BibitemShut {NoStop}%
\bibitem [{\citenamefont {Zhang}\ \emph {et~al.}(2021)\citenamefont {Zhang},
  \citenamefont {Gomes}, \citenamefont {Yao}, \citenamefont {Orth},\ and\
  \citenamefont {Iadecola}}]{Zhang.2021}%
  \BibitemOpen
  \bibfield  {author} {\bibinfo {author} {\bibfnamefont {F.}~\bibnamefont
  {Zhang}}, \bibinfo {author} {\bibfnamefont {N.}~\bibnamefont {Gomes}},
  \bibinfo {author} {\bibfnamefont {Y.}~\bibnamefont {Yao}}, \bibinfo {author}
  {\bibfnamefont {P.~P.}\ \bibnamefont {Orth}},\ and\ \bibinfo {author}
  {\bibfnamefont {T.}~\bibnamefont {Iadecola}},\ }\bibfield  {title} {\bibinfo
  {title} {{Adaptive variational quantum eigensolvers for highly excited
  states}},\ }\href {https://doi.org/10.1103/physrevb.104.075159} {\bibfield
  {journal} {\bibinfo  {journal} {Physical Review B}\ }\textbf {\bibinfo
  {volume} {104}},\ \bibinfo {pages} {075159} (\bibinfo {year} {2021})},\
  \Eprint {https://arxiv.org/abs/2104.12636} {2104.12636} \BibitemShut
  {NoStop}%
\bibitem [{\citenamefont {Zhang}\ \emph {et~al.}(2022)\citenamefont {Zhang},
  \citenamefont {Chen}, \citenamefont {Yuan},\ and\ \citenamefont
  {Yin}}]{Zhang.2022}%
  \BibitemOpen
  \bibfield  {author} {\bibinfo {author} {\bibfnamefont {D.-B.}\ \bibnamefont
  {Zhang}}, \bibinfo {author} {\bibfnamefont {B.-L.}\ \bibnamefont {Chen}},
  \bibinfo {author} {\bibfnamefont {Z.-H.}\ \bibnamefont {Yuan}},\ and\
  \bibinfo {author} {\bibfnamefont {T.}~\bibnamefont {Yin}},\ }\bibfield
  {title} {\bibinfo {title} {{Variational quantum eigensolvers by variance
  minimization}},\ }\href {https://doi.org/10.1088/1674-1056/ac8a8d} {\bibfield
   {journal} {\bibinfo  {journal} {Chinese Physics B}\ }\textbf {\bibinfo
  {volume} {31}},\ \bibinfo {pages} {120301} (\bibinfo {year}
  {2022})}\BibitemShut {NoStop}%
\bibitem [{\citenamefont {Hobday}\ \emph {et~al.}(2022)\citenamefont {Hobday},
  \citenamefont {Stevenson},\ and\ \citenamefont {Benstead}}]{Hobday.2022}%
  \BibitemOpen
  \bibfield  {author} {\bibinfo {author} {\bibfnamefont {I.}~\bibnamefont
  {Hobday}}, \bibinfo {author} {\bibfnamefont {P.}~\bibnamefont {Stevenson}},\
  and\ \bibinfo {author} {\bibfnamefont {J.}~\bibnamefont {Benstead}},\
  }\bibfield  {title} {\bibinfo {title} {{Variance minimisation on a quantum
  computer for nuclear structure}},\ }\bibfield  {journal} {\bibinfo  {journal}
  {arXiv}\ }\href {https://doi.org/10.48550/arxiv.2209.07820}
  {10.48550/arxiv.2209.07820} (\bibinfo {year} {2022}),\ \Eprint
  {https://arxiv.org/abs/2209.07820} {2209.07820} \BibitemShut {NoStop}%
\bibitem [{\citenamefont {Boyd}\ and\ \citenamefont
  {Koczor}(2022)}]{Boyd.2022}%
  \BibitemOpen
  \bibfield  {author} {\bibinfo {author} {\bibfnamefont {G.}~\bibnamefont
  {Boyd}}\ and\ \bibinfo {author} {\bibfnamefont {B.}~\bibnamefont {Koczor}},\
  }\bibfield  {title} {\bibinfo {title} {{Training Variational Quantum Circuits
  with CoVaR: Covariance Root Finding with Classical Shadows}},\ }\href
  {https://doi.org/10.1103/physrevx.12.041022} {\bibfield  {journal} {\bibinfo
  {journal} {Physical Review X}\ }\textbf {\bibinfo {volume} {12}},\ \bibinfo
  {pages} {041022} (\bibinfo {year} {2022})},\ \Eprint
  {https://arxiv.org/abs/2204.08494} {2204.08494} \BibitemShut {NoStop}%
\bibitem [{\citenamefont {Liu}\ \emph {et~al.}(2023)\citenamefont {Liu},
  \citenamefont {Zhang}, \citenamefont {Hsieh}, \citenamefont {Zhang},\ and\
  \citenamefont {Yao}}]{Liu.2023}%
  \BibitemOpen
  \bibfield  {author} {\bibinfo {author} {\bibfnamefont {S.}~\bibnamefont
  {Liu}}, \bibinfo {author} {\bibfnamefont {S.-X.}\ \bibnamefont {Zhang}},
  \bibinfo {author} {\bibfnamefont {C.-Y.}\ \bibnamefont {Hsieh}}, \bibinfo
  {author} {\bibfnamefont {S.}~\bibnamefont {Zhang}},\ and\ \bibinfo {author}
  {\bibfnamefont {H.}~\bibnamefont {Yao}},\ }\bibfield  {title} {\bibinfo
  {title} {{Probing many-body localization by excited-state variational quantum
  eigensolver}},\ }\href {https://doi.org/10.1103/physrevb.107.024204}
  {\bibfield  {journal} {\bibinfo  {journal} {Physical Review B}\ }\textbf
  {\bibinfo {volume} {107}},\ \bibinfo {pages} {024204} (\bibinfo {year}
  {2023})},\ \Eprint {https://arxiv.org/abs/2111.13719} {2111.13719}
  \BibitemShut {NoStop}%
\bibitem [{\citenamefont {Schlimgen}\ \emph {et~al.}(2021)\citenamefont
  {Schlimgen}, \citenamefont {Head-Marsden}, \citenamefont {Sager},
  \citenamefont {Narang},\ and\ \citenamefont {Mazziotti}}]{Schlimgen.2021}%
  \BibitemOpen
  \bibfield  {author} {\bibinfo {author} {\bibfnamefont {A.~W.}\ \bibnamefont
  {Schlimgen}}, \bibinfo {author} {\bibfnamefont {K.}~\bibnamefont
  {Head-Marsden}}, \bibinfo {author} {\bibfnamefont {L.~M.}\ \bibnamefont
  {Sager}}, \bibinfo {author} {\bibfnamefont {P.}~\bibnamefont {Narang}},\ and\
  \bibinfo {author} {\bibfnamefont {D.~A.}\ \bibnamefont {Mazziotti}},\
  }\bibfield  {title} {\bibinfo {title} {{Quantum Simulation of Open Quantum
  Systems Using a Unitary Decomposition of Operators}},\ }\href
  {https://doi.org/10.1103/physrevlett.127.270503} {\bibfield  {journal}
  {\bibinfo  {journal} {Physical Review Letters}\ }\textbf {\bibinfo {volume}
  {127}},\ \bibinfo {pages} {270503} (\bibinfo {year} {2021})},\ \Eprint
  {https://arxiv.org/abs/2106.12588} {2106.12588} \BibitemShut {NoStop}%
\bibitem [{\citenamefont {Seki}\ and\ \citenamefont
  {Yunoki}(2021)}]{Seki.2021}%
  \BibitemOpen
  \bibfield  {author} {\bibinfo {author} {\bibfnamefont {K.}~\bibnamefont
  {Seki}}\ and\ \bibinfo {author} {\bibfnamefont {S.}~\bibnamefont {Yunoki}},\
  }\bibfield  {title} {\bibinfo {title} {{Quantum Power Method by a
  Superposition of Time-Evolved States}},\ }\bibfield  {journal} {\bibinfo
  {journal} {PRX Quantum}\ }\textbf {\bibinfo {volume} {2}},\ \href
  {https://doi.org/10.1103/prxquantum.2.010333} {10.1103/prxquantum.2.010333}
  (\bibinfo {year} {2021}),\ \Eprint {https://arxiv.org/abs/2008.03661}
  {2008.03661} \BibitemShut {NoStop}%
\bibitem [{\citenamefont {Hehre}\ \emph {et~al.}(1969)\citenamefont {Hehre},
  \citenamefont {Stewart},\ and\ \citenamefont {Pople}}]{Hehre.1969}%
  \BibitemOpen
  \bibfield  {author} {\bibinfo {author} {\bibfnamefont {W.~J.}\ \bibnamefont
  {Hehre}}, \bibinfo {author} {\bibfnamefont {R.~F.}\ \bibnamefont {Stewart}},\
  and\ \bibinfo {author} {\bibfnamefont {J.~A.}\ \bibnamefont {Pople}},\
  }\bibfield  {title} {\bibinfo {title} {{Self‐Consistent Molecular‐Orbital
  Methods. I. Use of Gaussian Expansions of Slater‐Type Atomic Orbitals}},\
  }\href {https://doi.org/10.1063/1.1672392} {\bibfield  {journal} {\bibinfo
  {journal} {The Journal of Chemical Physics}\ }\textbf {\bibinfo {volume}
  {51}},\ \bibinfo {pages} {2657} (\bibinfo {year} {1969})}\BibitemShut
  {NoStop}%
\bibitem [{\citenamefont {RDMChem}(2023)}]{rdmchem}%
  \BibitemOpen
  \bibfield  {author} {\bibinfo {author} {\bibnamefont {RDMChem}},\ }\href@noop
  {} {\bibinfo {title} {{Quantum Chemistry Toolbox in Maple}}} (\bibinfo {year}
  {2023})\BibitemShut {NoStop}%
\bibitem [{\citenamefont {Liu}\ and\ \citenamefont {Nocedal}(1989)}]{Liu.1989}%
  \BibitemOpen
  \bibfield  {author} {\bibinfo {author} {\bibfnamefont {D.~C.}\ \bibnamefont
  {Liu}}\ and\ \bibinfo {author} {\bibfnamefont {J.}~\bibnamefont {Nocedal}},\
  }\bibfield  {title} {\bibinfo {title} {{On the limited memory BFGS method for
  large scale optimization}},\ }\href {https://doi.org/10.1007/bf01589116}
  {\bibfield  {journal} {\bibinfo  {journal} {Mathematical Programming}\
  }\textbf {\bibinfo {volume} {45}},\ \bibinfo {pages} {503} (\bibinfo {year}
  {1989})}\BibitemShut {NoStop}%
\end{thebibliography}%

\end{document}